\documentclass[preprint]{elsarticle}
\usepackage{graphicx}
\usepackage{multirow}
\usepackage{amsmath}
\usepackage{amssymb}
\usepackage{xspace}
\usepackage{gensymb}
\usepackage{booktabs}
\usepackage{hyperref}
\usepackage{aas_macros}
\usepackage{xcolor}
\usepackage{array}
\usepackage{adjustbox}
\usepackage{fancyhdr}

\usepackage{etoolbox}
\makeatletter
    \patchcmd{\tnotemark}{\ding{73}}{}{}{\@latex@error{Failed to path \string\tnotemark\space for \string\ding{73}}}
    \patchcmd{\tnotetext}{\ding{73}}{}{}{\@latex@error{Failed to path \string\tnotetext\space for \string\ding{73}}}

\newlength{\wdo}
\newcommand{\stroke}[1]{{$#1$}%
\settowidth{\wdo}{${#1}$} {\kern-\wdo}%
\partialvartstrokedint}

\makeatletter
\newcommand{\fancysep}{%
  \@afterindentfalse
  {\begin{center}
    \resizebox{0.8\linewidth}{0.4ex}{{%
        \fontsize{20}{24}\usefont{U}{webo}{xl}{n}{4}}}%
  \end{center}}\@afterheading}
\makeatother

{\vskip 1.0em
\framebox[\columnwidth][r]{%
\begin{minipage}[c]{\columnwidth}%
\vspace{-1.0em}%
#1%
\end{minipage}}}
{\vskip 1.0em}

\def\XXint#1#2#3{{\setbox0=\hbox{$#1{#2#3}{\int}$}
     \vcenter{\hbox{$#2#3$}}\kern-.5\wd0}}

\newcommand{\beq}{\begin{equation}}
\newcommand{\eeq}{\end{equation}}
\newcommand{\beqa}{\begin{eqnarray}}
\newcommand{\eeqa}{\end{eqnarray}}

\newcommand{\mpl}{\ensuremath{m_\textnormal{pl}}}

\newcommand{\burst}{{\sc burst}\xspace}

\newcommand{\gstar}{\ensuremath{g_\star}\xspace}

\newcommand{\gstars}{\ensuremath{g_{\star S}}\xspace}

\newcommand{\tcm}{\ensuremath{T_\textnormal{cm}}\xspace}

\newcommand{\ssamp}{\ensuremath{\langle|\mathcal{M}|^2\rangle}\xspace}

\newcommand{\nprates}{\ensuremath{n\leftrightarrow p}\xspace rates\xspace}

\newcommand{\nue}{{\ensuremath{\nu_{e}}}\xspace}

\newcommand{\bnue}{\ensuremath{\bar\nu_e}\xspace}

\newcommand{\spl}{\ensuremath{s_\textnormal{pl}}\xspace}

\newcommand{\ben}{\begin{enumerate}}
\newcommand{\een}{\end{enumerate}}

\newcommand{\twfo}{\ensuremath{T_\textnormal{WFO}}\xspace}
\newcommand{\mntau}{\ensuremath{\tau_n}\xspace}
\newcommand{\deltamnp}{\ensuremath{\delta m_{np}}\xspace}
\newcommand{\teq}{\ensuremath{T_\textnormal{eq}}\xspace}
\newcommand{\npratio}{\ensuremath{n/p}\xspace ratio\xspace}
\newcommand{\heiv}{\ensuremath{\,^4\textnormal{He}}\xspace}
\newcommand{\yp}{\ensuremath{Y_P}\xspace}

\begin{document}

\title{The surprising influence of late charged current weak interactions
on Big Bang Nucleosynthesis}


\author{E.~Grohs\corref{cor1}}
\ead{egrohs@umich.edu}
\address{Department of Physics, University of Michigan, Ann Arbor, Michigan
48109, USA}
\cortext[cor1]{Corresponding author}

\author{G.M.~Fuller\corref{cor2}}
\ead{gfuller@ucsd.edu}
\address{Department of Physics, University of California,
San Diego, La Jolla, California 92093, USA}

\date{\today}

\begin{abstract} 

The weak interaction charged current processes ($\nu_e+n\leftrightarrow p+e^-$;
$\bar\nu_e +p\leftrightarrow n+e^+$; $n\leftrightarrow p+e^-+\bar\nu_e$)
interconvert neutrons and protons in the early universe and have significant
influence on Big Bang Nucleosynthesis (BBN) light-element abundance yields,
particularly that for $^{4}{\rm He}$. We demonstrate that the influence of these
processes is still significant even when they operate well below temperatures
$T\sim0.7\,{\rm MeV}$ usually invoked for ``weak freeze-out,'' and in fact down
nearly into the alpha-particle formation epoch ($T \approx 0.1\,{\rm MeV}$).  This
physics is correctly captured in commonly used BBN codes, though this
late-time, low-temperature persistent effect of the isospin-changing weak
processes, and the sensitivity of the associated rates to lepton energy
distribution functions and blocking factors are not widely appreciated.  We
quantify this late-time influence by analyzing weak interaction rate dependence
on the neutron lifetime, lepton energy distribution functions, entropy, the
proton-neutron mass difference, and Hubble expansion rate.  The effects we
point out here render BBN a keen probe of any beyond-standard-model physics
that alters lepton number/energy distributions, even subtly, in epochs of the
early universe all the way down to near $T=100\,{\rm keV}$.

\end{abstract}




\maketitle

\section{Introduction}

%

In this paper we examine individual charged current weak interactions in the
Big Bang Nucleosynthesis (BBN) epoch and uncover a feature of these which is
surprising and explains how BBN can be sensitive to any physics which even
slightly alters neutrino energy distribution functions, even at relatively low
temperatures. The importance of this sensitivity for containing and probing
beyond-standard-model (BSM) physics  will be heightened as the precision of the
observationally inferred primordial abundances of deuterium
\cite{Cooke:2014do,2016MNRAS.455.1512C} and helium \cite{cmbs4uofc} increases
in the future.

BBN is a great success story of the marriage of nuclear and particle physics
and cosmology
\cite{1998RvMP...70..303S,2002PhR...370..333D,2012arXiv1208.0032S,2016RvMP...88a5004C}.
In essence, the light element abundances that emerge from the early universe
are set in a freeze-out from nuclear statistical equilibrium (NSE). NSE is
thermal and chemical equilibrium among all nuclei, where the abundance of any
nuclear species is given by a Saha equation that depends only on nuclear
binding energy, entropy-per-baryon, temperature, and the neutron-to-proton
ratio (i.e.\ the ratio of neutron to proton number densities; denoted \npratio)
\cite{B2FH}.  NSE obtains when the rates of strong and electromagnetic
nuclear reactions are fast compared to the rate at which thermodynamic
conditions change, in this case governed mostly by the Hubble expansion rate.
NSE freeze-out in the early universe is actually a series of freeze-outs of
individual nuclear reactions and is a protracted event, taking place over many
Hubble times. This endurance is a key reason why the eventual BBN light element
abundance yields are sensitive to slow and relatively small, weak
interaction-induced changes in the \npratio during the NSE freeze-out process.

The weakness of gravitation, along with an entropy-per-baryon
($\sim9\times{10}^9$ Boltzmann's constant per baryon) which is large on a
nuclear physics scale, and very small net electron, muon, and tau lepton
numbers, combine to dictate most of what happens in NSE freeze-out.
Gravitation drives the Hubble expansion, so it is inherently slow, and in the
relevant radiation-dominated conditions the Hubble expansion rate is $H\propto
T^2/m_{\rm pl}$, where $T$ is the plasma temperature and
$\mpl\simeq1.221\times10^{22}\,{\rm MeV}$ is the Planck mass.  This expansion
rate is so slow that at temperatures $T > 1\,{\rm MeV}$, the weak interaction
can maintain equilibrium by efficiently scattering and exchanging energy
between the seas of neutrinos and the photon-electron-positron plasma.  Typical
weak interaction rates involving neutrinos and electrons/positrons are $\sim
G_F^2\,T^5$, with $G_F\simeq1.166\times10^{-11}\,{\rm MeV}^{-2}$ the Fermi
coupling constant.  The number of baryons is very small (baryon-to-photon ratio
$\eta \approx 6\times {10}^{-10}$) and so lepton scattering and captures on
them are negligible in effecting energy transfer between the neutrino seas and
the photon-electron-positron plasma.  Nevertheless, the baryons and neutrino
seas are in chemical equilibrium at high temperature. As the universe expands
and the temperature drops, neutrinos cease to efficiently exchange energy with
the plasma and, eventually, completely decouple and free fall through
spacetime.  Likewise, the rates of neutron-to-proton converting (i.e.,
isospin-changing) weak interactions decrease as the temperature drops and,
eventually the \npratio will be dominated by neutron decay. The former
decoupling process is termed ``weak decoupling'', while the latter is dubbed
``Weak Freeze-Out'' (WFO).

The standard narrative has these as distinct events, weak decoupling at $T \sim
1\,{\rm MeV}$, and WFO at $T\sim 0.7\,{\rm MeV}$.  This narrative is incorrect
(see e.g., Refs.\ \cite{Dolgov:1997ne,neff:3.046,transport_paper}).

In fact, weak decoupling and WFO are not events that are instantaneous in time or
temperature. Weak decoupling, WFO, and aspects of NSE freeze-out, all occur
more or less contemporaneously over many, many Hubble times. This fact dictates
the use of a numerical approach to BBN, first done by Refs.\
\cite{Wagoner:1966pv,Wagoner:1969sy}, but it also sets up the sensitivity to
late-changing weak interactions and their dependence on neutrino distribution
functions which we examine in this paper.

The organization of this paper is as follows.  In section \ref{sec:nprates} we
give a brief exposition of the relevant charged current, isospin-changing weak
interaction processes and the rates of these in the BBN epoch.  In Sec.\
\ref{sec:results}, we present calculations which elucidate the physics of WFO
and highlight the inadequacy of neglecting {\it any} of the weak interactions
at temperatures $T\gtrsim100\,{\rm keV}$.  We give our conclusions in
Sec.\ \ref{sec:conclusion}.  Throughout this paper we use natural units,
$\hbar=c=k_B=1$, and assume neutrinos are massless at the temperature scales of
interest.

\section{The \nprates}
\label{sec:nprates}

Three processes convert neutrons ($n$) into protons ($p$) in the BBN epoch.
These are schematically shown as
\begin{align}
  \nue + n &\rightarrow p + e^-,\label{eq:np1}\\
  e^+ + n &\rightarrow p + \bnue,\label{eq:np2}\\
  n &\rightarrow p +\bnue+ e^-,\label{eq:np3}
\end{align}
where $e^\pm$ denote positron, electron, and $\nue,\,\bnue$ denote electron
neutrino, antineutrino.  Additionally, the three corresponding reverse processes
convert protons into neutrons
\begin{align}
  e^- + p &\rightarrow n + \nue,\label{eq:pn1}\\
  \bnue + p &\rightarrow n + e^+,\label{eq:pn2}\\
  \bnue + e^- + p &\rightarrow n.\label{eq:pn3}
\end{align}
Together, the rates associated with processes \eqref{eq:np1} -- \eqref{eq:pn3}
are the neutron-to-proton rates (\nprates).  At high temperature, the rates
associated with the capture processes in \eqref{eq:np1}, \eqref{eq:np2},
\eqref{eq:pn1}, and \eqref{eq:pn2} are all much faster than the Hubble
expansion rate.  Note processes \eqref{eq:np1} -- \eqref{eq:np2} have no energy
threshold for lepton capture because the neutron is heavier than the proton
with mass difference $\deltamnp\equiv m_n-m_p\approx1.293\,{\rm MeV}$.  In turn, the
lepton capture processes \eqref{eq:pn1} -- \eqref{eq:pn3} have an energy
threshold.  As we will see below, this threshold makes a significant difference
on the leverage these processes have on the \npratio, especially at low
temperature.  The three-body process \eqref{eq:pn3} is not a dominant
contribution.  Other weak interactions on nuclei are present during BBN
\cite{2010PhRvD..82l5017F}, but we focus on the \nprates for this work.

A self-consistent treatment of WFO includes calculating the \nprates within a
BBN nuclear-reaction network.  We use the \burst code to treat WFO
self-consistently with BBN \cite{GFKP-5pts:2014mn}, although other codes also
correctly capture the physics of WFO (see Refs.\
\cite{letsgoeu2,2008CoPhC.178..956P,2012CoPhC.183.1822A}).  To examine the
interplay between the \npratio and the primordial abundances, we will modify
the \nprates in \burst to employ limiting scenarios on WFO.  We expect the
altered scenarios to yield different results for the \npratio and abundances as
compared to baseline computations that do not employ the said limits.  Our aim
will be to assess how effective each scenario is for accurately characterizing
WFO and predicting primordial abundances.

To begin our analysis, we need expressions for the \nprates suitable for
computational implementation.  Ref.\ \cite{2016arXiv160509383B} gives a
prescription to calculate the collision integral for the full quantum kinetic
equation (QKE) for each process in \eqref{eq:np1} -- \eqref{eq:pn3}.  We do not
utilize the full QKE treatment in this work (see Refs.\
\cite{1991NuPhB.349..743B,AkhmedovBerezhiani,1993APh.....1..165R,2005PhRvD..71i3004S,
2007JPhG...34...47B,2013PhRvD..87k3010V,2013PrPNP..71..162B,
Gouvea,2014PhRvD..90l5040S,2015PhLB..747...27C}
for discussion on the QKEs).  Instead, we employ multiple approximations to
facilitate ease in numerically computing the \nprates.  We start with the
collision integral for the QKE in processes \eqref{eq:np1} and \eqref{eq:pn1},
i.e.\ the gain and loss terms in the reaction $\nue n\leftrightarrow p\,e^-$.
Ref.\ \cite{2016arXiv160509383B} writes the collision term for the change in
neutrino occupation number, i.e.\ $df_{\nue}/dt$.  The change in the total
number of neutrinos is identical to the change in the total number of neutrons,
implying we can write an integrated collision term with $n(Q_1)+
\nu_e(Q_2)\leftrightarrow p(Q_3)+e^-(Q_4)$, where $Q_i$ is the four-momentum
for particle $i$, for the change in the number of neutrons.  The result is the
following Boltzmann equation \cite{Dodelson:2003mc}
\begin{align}
  \frac{dY_n}{dt}\biggr|_{\nue n\leftrightarrow pe^-} = \frac{1}{n_b}\int
  &\widetilde{dq_1}\widetilde{dq_2}\widetilde{dq_3}\widetilde{dq_4}\nonumber\\
  \times&(2\pi)^4\delta^{(4)}(Q_1 + Q_2 - Q_3 - Q_4)\nonumber\\
  \times&\ssamp F(E_1,E_2,E_3,E_4),\label{eq:boltz1}
\end{align}
where $Y_n$ is the neutron abundance.  We define abundances to be the ratio of
number densities, $Y_i\equiv n_i/n_b$, where $n_i$ is the number density of
species $i$ and $n_b$ is the baryon number density.  In Eq.\ \eqref{eq:boltz1},
we use the phase-space notation
\beq
  \widetilde{dq_i}\equiv\frac{d^3q_i}{(2\pi)^32E_i},
\eeq
for three-momentum magnitude $q_i$ and energy $E_i$.
$\delta^{(4)}(Q_1+Q_2-Q_3-Q_4)$ is a four-momentum conserving Dirac delta
function and $F$ is the statistical factor appropriate for Fermi-Dirac
statistics
\begin{alignat}{3}
  F = &&f_p(E_3)f_{e^-}(E_4)[1-f_n(E_1)][1-f_{\nue}(E_2)]\nonumber\\
  &-&f_n(E_1)f_{\nue}(E_2)[1-f_p(E_3)][1-f_{e^-}(E_4)]\label{eq:stat_fact},
\end{alignat}
where $f_i(E_i)$ is the occupation number for species $i$ at energy $E_i$.  We
depart from the quantum description of Ref.\ \cite{2016arXiv160509383B} and use
neutrino occupation numbers instead of the generalized density matrices for
this work.  For the specific processes in Eqs.\ \eqref{eq:np1} and
\eqref{eq:pn1}, the summed-squared-amplitude is \cite{2016arXiv160509383B}
\begin{alignat}{3}
  \ssamp=2^4G_F^2\{&&(g_A+1)^2(Q_1\cdot Q_2)(Q_3\cdot Q_4)\nonumber\\
  &+&(g_A-1)^2(Q_1\cdot Q_4)(Q_2\cdot Q_3)\nonumber\\
  &+&(g_A^2-1)m_nm_p(Q_2\cdot Q_4)\},\label{eq:ssamp}
\end{alignat}
where $g_A\simeq1.27$ is the axial-vector coupling and $m_n$ ($m_p$) is the
neutron (proton) vacuum mass. We can simplify Eq.\ \eqref{eq:ssamp}, if we make
the following approximations:
\ben
  \item Assume the energy is much larger than the magnitude of
  the three momenta for the baryons. 
  \item Assume the mass of the baryons is approximately equal to the energy.
  \item Assume the neutrino is massless, i.e.\ $E_2=q_2$.
\een
Under the above assumptions, Eq.\ \eqref{eq:ssamp} simplifies to:
\beq
  \ssamp = 2^4G_F^2E_1E_2E_3E_4
  \left[1+3g_A^2 + (g_A^2-1)\frac{q_4}{E_4}\cos\theta_{24}\right],\label{eq:ssamp2}
\eeq
where $\theta_{24}$ is the angle between $\vec{q}_2$ and $\vec{q}_4$.  We are
free to define our coordinate system such that $\theta_{24}$ is coincident with
one of the elevation angles in either $d^3q_2$ or $d^3q_4$.  Therefore, the
integral over the $\cos\theta_{24}$ term in Eq.\ \eqref{eq:ssamp2} will vanish,
leaving a remaining nonzero portion of the summed-squared-amplitude
\beq
  \ssamp \rightarrow 2^4G_F^2E_1E_2E_3E_4(1+3g_A^2).\label{eq:ssamp3}
\eeq

Eq.\ \eqref{eq:ssamp3} contains no angular dependence and is straightforward to
integrate in Eq.\ \eqref{eq:boltz1} with additional approximations.  We
use Fermi-Dirac (FD) statistics to describe the \nue and $e^-$, but simplify to
Maxwell-Boltzmann (MB) statistics for the neutron and proton.  This
simplification allows us to make the approximation $[1-f_{n,p}]\approx 1$, which
in turn simplifies the statistical factor $F$.  We relabel the second energy
dummy variable, $E_2\rightarrow E_\nu$, and reduce Eq.\ \eqref{eq:boltz1} to a
single integration
\begin{align}
  \frac{dY_n}{dt}\biggr|_{\nue n\leftrightarrow pe^-}
  = \frac{G_F^2(1+3g_A^2)}{2\pi^3}\int\limits_0^{\infty}dE_\nu
  &\,E_\nu^2(E_\nu+\deltamnp)\sqrt{(E_\nu+\deltamnp)^2-m_e^2}\nonumber\\
  \times&\{Y_p[1-f_{\nue}(E_\nu)][f_{e^-}(E_\nu+\deltamnp)]\nonumber\\
  &-Y_n[f_{\nue}(E_\nu)][1-f_{e^-}(E_\nu+\deltamnp)]\},\label{eq:boltz2}
\end{align}
where $m_e$ is the electron rest mass.  Eq.\ \eqref{eq:boltz2} contains two
contributions: a creation component proportional to the proton abundance $Y_p$,
deduced from process \eqref{eq:pn1}; and a destruction component proportional
to the neutron abundance $Y_n$, deduced from \eqref{eq:np1}.  We write the
creation component as
\beq
  \left(\frac{dY_n}{dt}\right)^+
  \equiv \frac{G_F^2(1+3g_A^2)}{2\pi^3}Y_p \tcm^5\mathcal{N},\label{eq:dydtplus}
\eeq
where we have written the integral over $E_\nu$ as $\tcm^5\mathcal{N}$.  \tcm
is the comoving temperature parameter and is inversely proportional to the
scale factor, $a$, such that $\tcm=T_{\rm in}(a_{\rm in}/a)$ where $T_{\rm in}$
($a_{\rm in}$) is the plasma temperature (scale factor) taken at an initial
epoch of our choosing \cite{transport_paper}.  For the purposes of studying
WFO, we will set $T_{\rm in}=10\,{\rm MeV}$.  $\mathcal{N}$ is a dimensionless
number and depends on the temperature quantities through the ratios
$\deltamnp/\tcm$, $m_e/\tcm$, and $\tcm/T$.

We can approximate the Hubble expansion rate during radiation dominated
conditions using the effective degrees of freedom statistic, \gstar, to yield
\cite{1990eaun.book.....K}
\beq
  H = \sqrt{\frac{8\pi}{3\mpl^2}\,\frac{\pi^2}{30}\gstar \tcm^4},\label{eq:hub}
\eeq
If we equate Eq.\ \eqref{eq:hub} with Eq.\ \eqref{eq:dydtplus}, we find a
comoving temperature value, \teq, where the two rates are equal
\beq
  \teq \equiv \left[\frac{2\pi^3}{\mpl G_F^2(1+3g_A^2)}
  \frac{1}{Y_p\mathcal{N}}\sqrt{\frac{8\pi^3}{90}\gstar}\right]^{1/3}.
\eeq
If we take $\mathcal{N}\simeq4!$, $\gstar=10.75$, and $Y_p\simeq1$, we
calculate $\teq \simeq0.9\,{\rm MeV}$ in line with the nominal estimate of
$0.7\,{\rm MeV}$ for WFO [see Eq.\ (57) in Ref.\ \cite{2002PhR...370..333D} and
references therein].  For $\tcm>\teq$, the creation term $(dY_n/dt)^+$ is
larger than the Hubble rate, and vice versa for $\tcm<\teq$.

In our calculations, we assume neutrinos instantaneously decoupled from the
plasma at a comoving temperature parameter $\tcm>\teq$.  The nucleosynthesis
effects from the altered WFO scenarios dwarf the contributions from neutrino
energy transport (see Refs.\ \cite{Dolgov:1997ne,neff:3.046,transport_paper}).
Furthermore, we do not include higher order corrections to the \nprates from
finite temperature radiative corrections or modified dispersion relations of
the charged leptons.  See Refs.\
\cite{1982PhRvD..26.2694D,1982NuPhB.209..372C,1999PhRvD..59j3502L} for detailed
accounting of the changes to the primordial abundances induced from these
corrections.  We do include Coulomb corrections from Refs.\
\cite{1980ApJS...42..447F,2010PhRvD..81f5027S} to processes \eqref{eq:np1},
\eqref{eq:np3}, \eqref{eq:pn1}, and \eqref{eq:pn3}.  In addition, we include
the zero-temperature radiative corrections of Refs.\
\cite{1982PhRvD..26.2694D,1999PhRvD..59j3502L} to all six processes.

In simplifying the amplitude in Eq.\ \eqref{eq:ssamp3}, we treated the baryons
as nonrelativistic when we set $m\sim E$.  This allowed us to separate any
dependence the energy-conserving part of the four-dimensional delta function
had on the nucleon energies [$E_1$ and $E_3$ in Eq.\ \eqref{eq:boltz1}],
thereby eliminating the explicit presence of the baryon MB occupation numbers.
The approximation of large baryon rest mass employed here induces a change of a
few percent.  See Refs.\
\cite{2016arXiv160509383B,1999PhRvD..59j3502L,1997PhRvD..56.3191L} for details
on how to construct the Boltzmann equation with baryons obeying MB or FD
statistics without the large baryon mass approximation.  We neglect these
corrections here in the spirit of exploring the leverage that weak
isospin-changing interactions have during the BBN epoch.  However, our
conclusion that these processes are surprisingly effective at low temperature,
highlight the need for including small corrections in a high-precision
assessment of BBN absolute abundance yields.

Reference \cite{2016arXiv160509383B} further gives the prescription to find \ssamp
for the other two reactions, namely $e^+n\leftrightarrow p\,\bnue$ and
$n\leftrightarrow p\,\bnue e^-$.  Using the same approximations in determining
Eq.\ \eqref{eq:ssamp3} for $\nue n\leftrightarrow pe^-$, the expressions for
\ssamp in the two other reactions are identical to that in \eqref{eq:ssamp3}.
In order to account for the contribution of each rate to either the neutron or
proton abundance change, we will write the Boltzmann equations as the following
\begin{align}
  \frac{dY_n}{dt}=&-Y_n(\lambda_{\nue n} + \lambda_{e^+n} + \lambda_{n\,{\rm decay}})
  +Y_p(\lambda_{e^-p} + \lambda_{\bnue p} + \lambda_{\bnue e^-p}),\label{eq:dyndt}\\
  \frac{dY_p}{dt}=&\,Y_n(\lambda_{\nue n} + \lambda_{e^+n} + \lambda_{n\,{\rm decay}})
  -Y_p(\lambda_{e^-p} + \lambda_{\bnue p} + \lambda_{\bnue e^-p}).\label{eq:dypdt}
\end{align}
The neutron destruction (proton creation) rates are $\lambda_{\nue n},
\lambda_{e^+n}, \lambda_{n\,{\rm decay}}$ and correspond to processes
\eqref{eq:np1} -- \eqref{eq:np3}, respectively.  The neutron creation (proton
destruction) rates are $\lambda_{e^-p}, \lambda_{\bnue p}, \lambda_{\bnue
e^-p}$ and correspond to processes \eqref{eq:pn1} -- \eqref{eq:pn3},
respectively.  Eqs.\ \eqref{eq:dyndt} and \eqref{eq:dypdt} give the changes in
the neutron and proton abundances from the weak interactions.  Nuclear
reactions such as deuterium ($d$) production in $n(p,\gamma)d$ also affect the
free neutron and proton abundances and would enter into both expressions
\eqref{eq:dyndt} and \eqref{eq:dypdt}.  For the sake of brevity, we have
ignored the nuclear contributions in writing Eqs.\ \eqref{eq:dyndt} and
\eqref{eq:dypdt} so that we can define the weak rates $\lambda_i$ corresponding
to processes \eqref{eq:np1} -- \eqref{eq:pn3}.  In the actual numerical
calculations, the nuclear reaction contributions are indispensable in Eqs.\
\eqref{eq:dyndt} and \eqref{eq:dypdt}, and are compulsory for baryon number
conservation.

We arrive at the following expressions for the six \nprates in Eqs.\
\eqref{eq:dyndt} and \eqref{eq:dypdt}
\begin{align}
  \lambda_{\nue n} = \frac{G_F^2(1+3g_A^2)}{2\pi^3}\int\limits_0^{\infty}dE_\nu\,
  &C(E_\nu+\deltamnp)Z(E_\nu+\deltamnp,E_\nu)\nonumber\\
  \times& E_\nu^2(E_\nu+\deltamnp)\sqrt{(E_\nu+\deltamnp)^2-m_e^2}\nonumber\\
  \times&[f_{\nue}(E_\nu)][1-f_{e^-}(E_\nu+\deltamnp)],\label{eq:urcap1}
\end{align}
\begin{align}
  \lambda_{e^+n} = \frac{G_F^2(1+3g_A^2)}{2\pi^3}\int\limits_{\deltamnp+m_e}^{\infty}dE_\nu\,
  &Z(E_\nu-\deltamnp,E_\nu)\nonumber\\
  \times& E_\nu^2(E_\nu-\deltamnp)\sqrt{(E_\nu-\deltamnp)^2-m_e^2}\nonumber\\
  \times&[1-f_{\bnue}(E_\nu)][f_{e^+}(E_\nu-\deltamnp)],\label{eq:urcap2}
\end{align}
\begin{align}
  \lambda_{n\,{\rm decay}} = \frac{G_F^2(1+3g_A^2)}{2\pi^3}\int\limits_{0}^{\deltamnp-m_e}dE_\nu\,
  &C(\deltamnp-E_\nu)Z(\deltamnp-E_\nu,E_\nu)\nonumber\\
  \times&E_\nu^2(\deltamnp-E_\nu)\sqrt{(\deltamnp-E_\nu)^2-m_e^2}\nonumber\\
  \times&[1-f_{\bnue}(E_\nu)][1-f_{e^-}(\deltamnp-E_\nu)],\label{eq:ndecay}
\end{align}
\begin{align}
  \lambda_{e^-p} = \frac{G_F^2(1+3g_A^2)}{2\pi^3}\int\limits_0^{\infty}dE_\nu\,
  &C(E_\nu+\deltamnp)Z(E_\nu+\deltamnp,E_\nu)\nonumber\\
  \times& E_\nu^2(E_\nu+\deltamnp)\sqrt{(E_\nu+\deltamnp)^2-m_e^2}\nonumber\\
  \times&[1-f_{\nue}(E_\nu)][f_{e^-}(E_\nu+\deltamnp)],\label{eq:urcan1}
\end{align}
\begin{align}
  \lambda_{\bnue p} = \frac{G_F^2(1+3g_A^2)}{2\pi^3}\int\limits_{\deltamnp+m_e}^{\infty}dE_\nu\,
  &Z(E_\nu-\deltamnp,E_\nu)\nonumber\\
  \times& E_\nu^2(E_\nu-\deltamnp)\sqrt{(E_\nu-\deltamnp)^2-m_e^2}\nonumber\\
  \times&[f_{\bnue}(E_\nu)][1-f_{e^+}(E_\nu-\deltamnp)],\label{eq:urcan2}
\end{align}
\begin{align}
  \lambda_{\bnue e^-p} = \frac{G_F^2(1+3g_A^2)}{2\pi^3}\int\limits_{0}^{\deltamnp-m_e}dE_\nu\,
  &C(\deltamnp-E_\nu)Z(\deltamnp-E_\nu,E_\nu)\nonumber\\
  \times& E_\nu^2(\deltamnp-E_\nu)\sqrt{(\deltamnp-E_\nu)^2-m_e^2}\nonumber\\
  \times&[f_{\bnue}(E_\nu)][f_{e^-}(\deltamnp-E_\nu)].\label{eq:indecay}
\end{align}
$C(E_e)$ and $Z(E_e,E_\nu)$ are modifications to the integrand from Coulomb
corrections and zero-temperature radiative corrections, respectively, for
charged-lepton-energy argument $E_e$.  The quantities $f_i$ give the lepton
occupation numbers for $i=\nue,\bnue, e^-, e^+$.  In this work, the occupation
numbers are always assumed to be FD in character
\beq\label{eq:fd}
  f_i^{\rm (FD)}(E;T_i,\mu_i) = \frac{1}{e^{(E-\mu_i)/T_i} + 1}.
\eeq
We use \tcm for the temperature quantity in the neutrino occupation numbers,
and $T$ for the plasma temperature in the charged lepton occupation numbers.
In addition, we evolve the electron chemical potential, $\mu_e$, for use in the
electron occupation numbers.  The positron chemical potential is always assumed
to be equal in magnitude and opposite in sign of the electron chemical
potential.  We take the electron neutrino and antineutrino chemical potentials
to be zero.  Our expressions for the \nprates in Eqs.\ \eqref{eq:urcap1} --
\eqref{eq:indecay} are identical to those of Ref.\ \cite{2009PhRvD..79j5001S}
except they contain the zero-temperature radiative corrections of Ref.\
\cite{1982PhRvD..26.2694D}, a different notation for the weak coefficient in
front of the integral, and some corrections to typographical errors.

We used the same set of approximations to simplify \ssamp present in Eqs.\
\eqref{eq:urcap1} -- \eqref{eq:indecay}. The result of this set of
approximations is that the same coefficient is in front of the integral over
$E_\nu$, namely $G_F^2(1+3g_A^2)/2\pi^3$, for all six rates in Eqs.\
\eqref{eq:urcap1} -- \eqref{eq:indecay}.  We do not explicitly calculate this
coefficient in our code using currently accepted values of $G_F$ and $g_A$.  As
an alternative, we can solve for the coefficient using the free neutron mean
lifetime in vacuum, $\tau_n$, which we simply refer to as the neutron lifetime.  
The expression for the vacuum decay rate of free neutrons is
\begin{align}
  \frac{1}{\tau_n} = \frac{G_F^2(1+3g_A^2)}{2\pi^3}
  \int\limits_{0}^{\deltamnp-m_e}dE_\nu\,
  &C(\deltamnp-E_\nu)Z(\deltamnp-E_\nu,E_\nu)\nonumber\\
  \times&E_\nu^2(\deltamnp-E_\nu)\sqrt{(\deltamnp-E_\nu)^2-m_e^2},\label{eq:ndecay_vac}
\end{align}
Eq.\ \eqref{eq:ndecay_vac} is identical to Eq.\ \eqref{eq:ndecay} with
$f_{\bnue,e^-}(E)=0$.  We can solve Eq.\ \eqref{eq:ndecay_vac} for the
coefficient $G_F^2(1+3g_A^2)/(2\pi^3)$ in terms of $\tau_n$ and a number for
the phase-space integral.  Ref.\ \cite{2012PhRvC..85f5503S} gives the neutron
lifetime as $\tau_n=882.5\pm2.1\,{\rm s}$, and Ref.\ \cite{2013PhRvL.111v2501Y}
gives $887.7\pm3.1\,{\rm s}$.  As an average, we will take
$\langle\tau_n\rangle=885.1\,{\rm s}$.

\section{Results}
\label{sec:results}

\begin{figure}
  \includegraphics[width=\columnwidth]{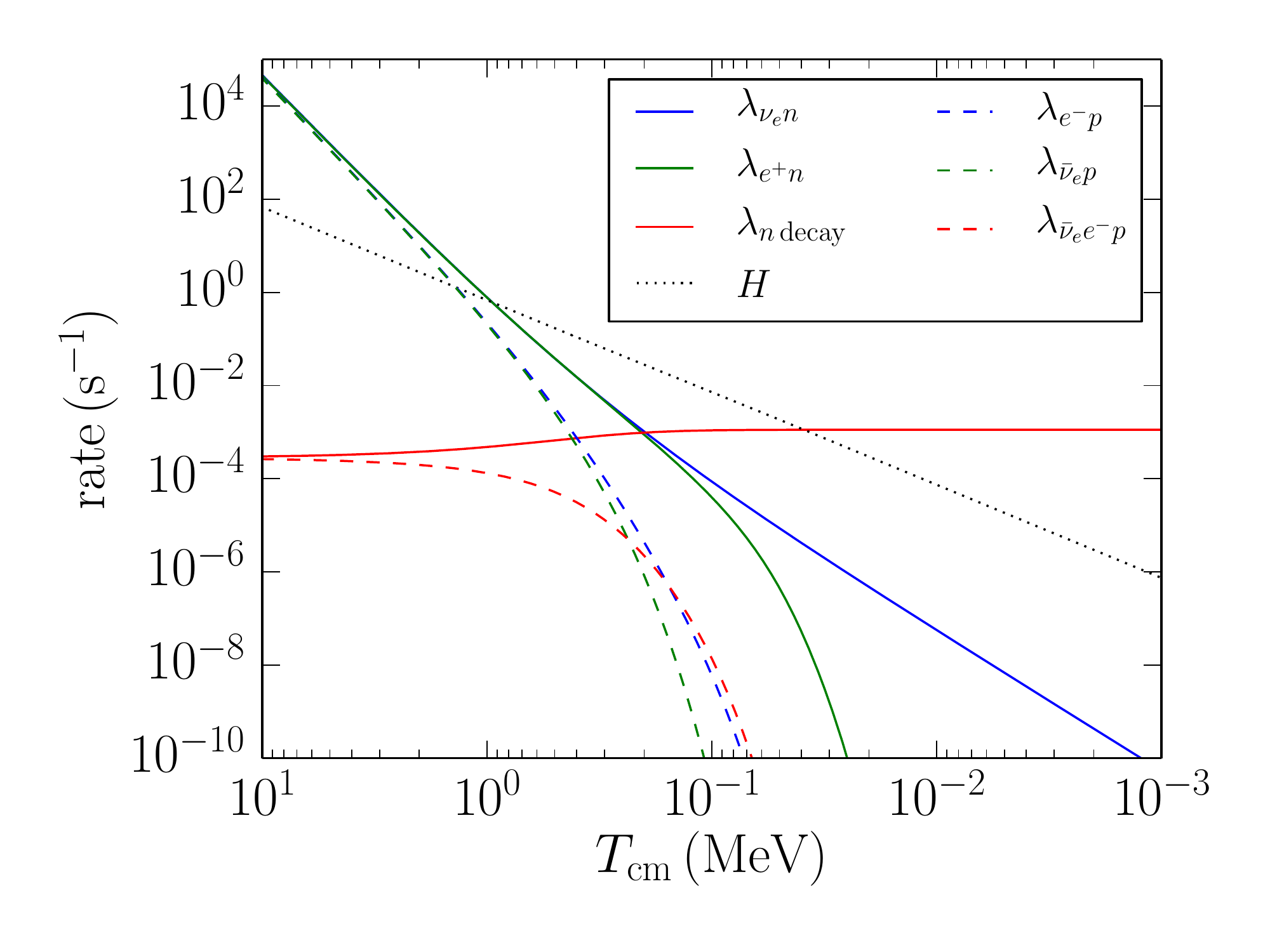}
  \caption{\label{fig:plot_rates_tcm} Rates as a function of comoving
  temperature parameter \tcm.  The \nprates, $\lambda$, include all six
  processes in \eqref{eq:np1} -- \eqref{eq:pn3}.  Also plotted is the Hubble
  rate, $H$, for comparison.  Our adopted neutron lifetime is $\tau_n=885.1\,{\rm s}$.}
\end{figure}

Figure \ref{fig:plot_rates_tcm} shows a plot of each of the six rates in Eqs.\
\eqref{eq:urcap1} -- \eqref{eq:indecay} versus \tcm. For comparison purposes,
the Hubble expansion rate ($H$) is also plotted.  The solid (dashed) curves
gives the rates for the neutron destruction (creation) processes.  In all three
reactions, the rate of neutron destruction is larger than that for creation.
We can determine the difference in strength between neutron destruction and
creation if we examine the explicit expressions in Eqs.\ \eqref{eq:urcap1} --
\eqref{eq:indecay}.  Consider first the rates for free neutron decay and the
inverse three-body process, Eqs.\ \eqref{eq:ndecay} and \eqref{eq:indecay}
respectively.  The equations are identical except for the expressions with the
FD occupation numbers.  Eq.\ \eqref{eq:indecay} is proportional to the
occupation numbers $[f_{\bnue,e^-}]$ directly, as opposed to Eq.\
\eqref{eq:ndecay} which is proportional to the blocking factors
$[1-f_{\bnue,e^-}]$.  For zero chemical potential, which is the case for the
neutrinos, $[1-f(E)]\ge f(E)$ for all values of $E$ if we assume FD occupation
numbers as in Eq.\ \eqref{eq:fd}.  Therefore, the rate for free neutron decay
is enhanced compared to the rate for the three-body process with regard to the
neutrino occupation number contribution.  The previous argument does not
readily apply to the electron occupation numbers because the electron sea has a
positive chemical potential.  However, the chemical potential is on order the
baryon number multiplied by the temperature at early times.  The argument in
the exponential of the occupation number is $(E-\mu_e)/T\approx E/T - 10^{-9}$
implying that for energies $E>10^{-9}T$, the blocking factor is larger than the
stand-alone occupation number.  Therefore, the electron and electron
antineutrino occupation numbers give rise to the difference in strength between
the two rates: $\lambda_{n\,{\rm decay}}\ge\lambda_{\bnue e^-p}$ at all
temperatures in Fig.\ \ref{fig:plot_rates_tcm}.

The above argument for the difference in strength of the $n\leftrightarrow
p\,\bnue e^-$ rates is not immediately applicable to the two capture reactions,
$\nue n\leftrightarrow p\,e^-$ and $e^+n\leftrightarrow p\,\bnue$.  There is a
reactant and product lepton in all four capture processes, so we cannot
directly compare $f_if_j$ to $[1-f_i][1-f_j]$ like we did in Eqs.\
\eqref{eq:ndecay} and \eqref{eq:indecay}.  The key to understanding the
comparative strength of the neutron destruction over creation rates is the
threshold energy, $E_{\rm thresh}$, absent in the destruction channel but
present in the creation channel.  The threshold energies are $E_{\rm
thresh}^{(e)}=\deltamnp$ for the electron in process \eqref{eq:pn1} and $E_{\rm
thresh}^{(\nu)}=\deltamnp+m_e$ for the electron antineutrino in process
\eqref{eq:pn2}.  In either capture reaction, $f(E\sim E_{\rm thresh})$ is
strictly less than $[1-f(E\sim E_{\rm thresh})]$ for the lepton requiring a
threshold energy.  For the lepton which does not require a threshold energy
(\nue in \eqref{eq:np1} and $e^+$ in \eqref{eq:np2}),
$f(E\sim0)\lesssim[1-f(E\sim0)]$.  The threshold energy reduces the allowed
phase-space for the process to occur.  At high temperatures, the threshold is
negligible and the reduction in phase space minimal. The neutron creation rate
is equal to the destruction rate as seen in the high-temperature range of Fig.\
\ref{fig:plot_rates_tcm}.  At lower temperatures, the rates begin to diverge
once the threshold becomes significant.

The neutron destruction rates are larger than the corresponding creation rates
for either the capture reactions or free neutron decay.  The result is a
decrease in the neutron abundance.  Standard BBN produces mostly $^1{\rm H}$
and \heiv.  The neutrons which survive WFO will become bound in the \heiv
isotope.  Therefore, when investigating altered scenarios of WFO, the principal
cosmological observable we will monitor is the primordial mass fraction of
$^4{\rm He}$, namely $Y_P\equiv4Y_{^4{\rm He}}$ and not to be confused with the
free proton abundance $Y_p$.  We use our adopted value of the neutron lifetime,
$\tau_n=885.1\,{\rm s}$, to obtain a $Y_P$ baseline of
\beq
  Y_P^{\rm (base)}[\mntau=885.1\,{\rm s}] = 0.2477.\label{eq:yp_baseline}
\eeq

\subsection{First scenario}

\begin{figure}
  \includegraphics[width=\columnwidth]{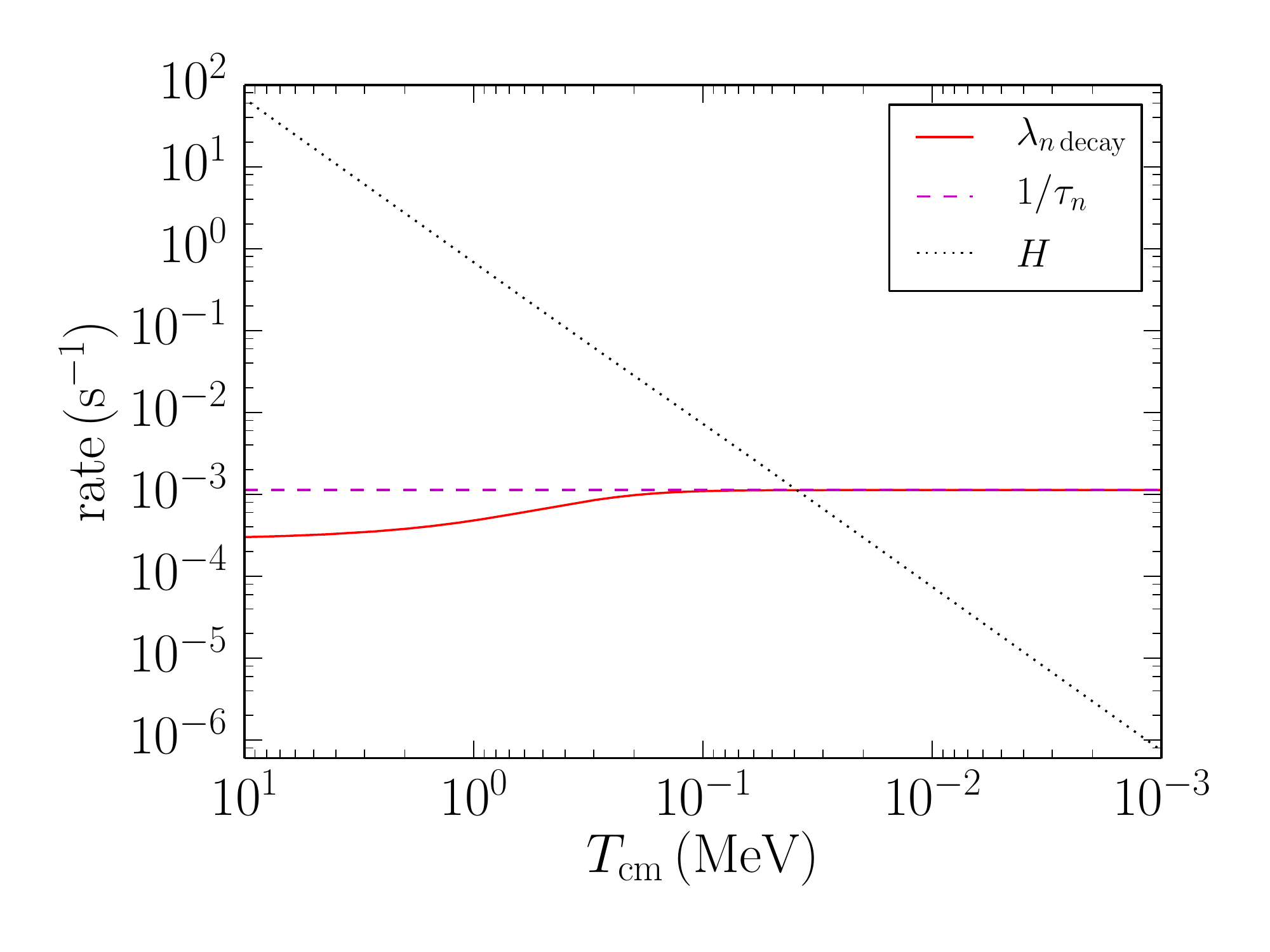}
  \caption{\label{fig:plot_ndecay_tcm}The in-medium free neutron decay rate,
  $\lambda_{n\,{\rm decay}}$, as a function of the comoving temperature parameter
  \tcm.  Also plotted is the Hubble expansion rate, $H$, and the vacuum neutron
  decay rate, $1/\tau_n$, for comparison.  Note that phase-space blocking
  decreases $\lambda_{n\,{\rm decay}}$ relative to the vacuum neutron decay rate
  at temperatures $\tcm\gtrsim100\,{\rm keV}$.  Our adopted neutron lifetime is
  $\mntau=885.1\,{\rm s}$.}
\end{figure}

In Fig.\ \ref{fig:plot_rates_tcm}, $\lambda_{n\,{\rm decay}}$ is not constant
with \tcm.  In medium, free neutron decay is inhibited by the FD blocking
factors $[1-f_{\bnue}]$ and $[1-f_{e^-}]$.  Blocking reduces $\lambda_{n\,{\rm
decay}}$, as shown in Fig.\ \ref{fig:plot_ndecay_tcm}.  In vacuum we drop the
FD blocking factors [like we did in Eq.\ \eqref{eq:ndecay_vac}], but in medium
we must include them.  At high temperatures, the FD occupation numbers for both
$e^-$ and \bnue are $\sim0.5$ for the range of energy values $0<E<\deltamnp$.
Therefore, the FD blocking factors are each $\sim0.5$ and the overall rate for
$\lambda_{n\,{\rm decay}}$ is $\sim0.25$ the vacuum rate.  This is evident in
Fig.\ \ref{fig:plot_ndecay_tcm} where the value of $\lambda_{n\,{\rm decay}}$
at high temperature is a factor of four lower than the value at low
temperature.  Eventually, the free neutron decay rate converges to the vacuum
rate at $\tcm\sim100\,{\rm keV}$, implying that FD blocking is negligible at
this epoch.  However, \heiv formation occurs during this $\tcm\sim100\,{\rm
keV}$ epoch.  In-medium free neutron decay is the dominant weak
isospin-changing process before \heiv nuclei form, as shown in Fig.\
\ref{fig:plot_rates_tcm}.  Modifications to the free neutron decay rate at late
epochs, $300\,{\rm keV}\gtrsim\tcm\gtrsim100\,{\rm keV}$, will alter the
\npratio and produce changes in \yp.  If $\lambda_{n\,{\rm decay}}\le1/\tau_n$,
then we would expect more neutrons to survive into the BBN epoch.  The net
result should be an increase in $Y_P$ compared to the scenario
$\lambda_{n\,{\rm decay}}=1/\tau_n$.

The first scenario of altered WFO we investigate is the sensitivity of \yp to
in-medium free neutron decay and the role of phase-space blocking factors.  Our
procedure is as follows for this scenario.  At high temperature we calculate
all six of the \nprates using the expressions in Eqs.\ \eqref{eq:urcap1} --
\eqref{eq:indecay}.  Once \tcm reaches a user-specified temperature, we no
longer calculate the free neutron decay rate using \eqref{eq:ndecay}.  At and
below the specified temperature, we set the free neutron decay rate equal to
the vacuum decay rate, namely $\lambda_{n\,{\rm decay}}\rightarrow1/\mntau$, in
essence ignoring the FD blocking factors $[1-f_{\bnue,e^-}]$ in Eq.\
\eqref{eq:ndecay}.  We include and calculate the three-body rate,
$\lambda_{\bnue e^-p}$, at all temperatures.  Excluding the three-body rate
produces a relative change of $\sim1.5\times10^{-3}$ in \yp and has little
effect for $\tcm\lesssim300\,{\rm keV}$.  Within our code, we implement the
following rate expressions, denoted $\lambda^{\rm (Sc.\,1)}$ for this first
scenario \begin{align} \lambda_{n\,{\rm decay}}^{\rm (Sc.\,1)}(\tcm)=&
\begin{cases} \lambda_{n\,{\rm decay}}&{\rm if}\quad\tcm>\twfo\\ 1/\mntau&{\rm
if}\quad\tcm<\twfo \end{cases},\nonumber\\ \lambda_i^{\rm
(Sc.\,1)}(\tcm)=&\quad\lambda_i, \end{align} where $i=\nue
n,\,e^+n,\,e^-p,\,\bnue p,\,\bnue e^-p$ for the rates in Eqs.\
\eqref{eq:urcap1}, \eqref{eq:urcap2}, \eqref{eq:urcan1}, \eqref{eq:urcan2},
\eqref{eq:indecay}.  \twfo is the user-specified ``weak freeze-out
temperature.''

\begin{figure}
  \includegraphics[width=\columnwidth]{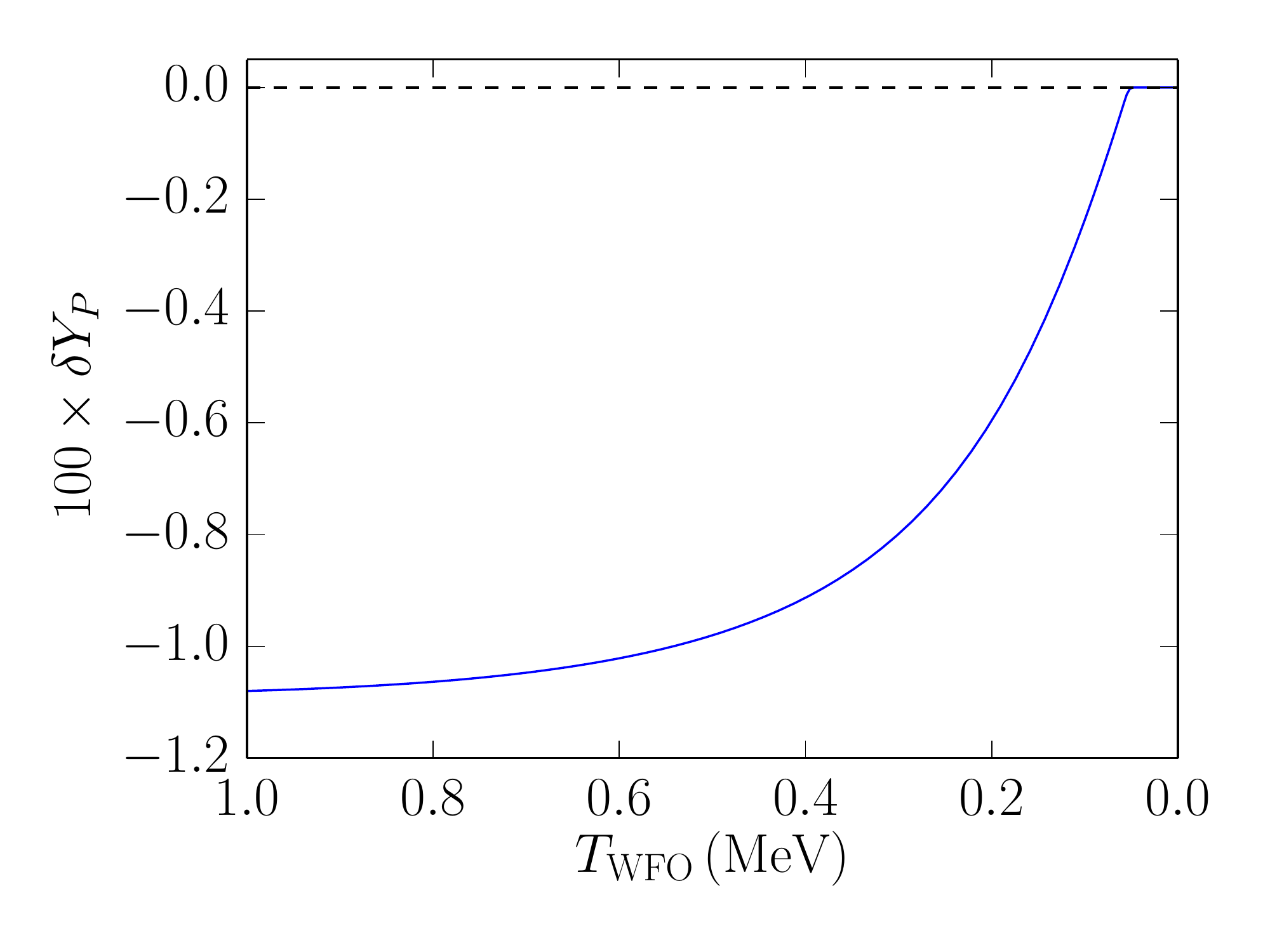}
  \caption{\label{fig:plot_scenario1}The relative difference in $Y_P$ compared
  to the baseline [Eq.\ \eqref{eq:deltayp_885}] as a function of \twfo in the
  first scenario of altered WFO. All six processes utilize the expressions in
  Eqs.\ \eqref{eq:urcap1} -- \eqref{eq:indecay} until \tcm reaches the
  user-specified \twfo.  For $\tcm<\twfo$, the free neutron decay rate is set to
  the vacuum rate, $1/\mntau$. Our adopted neutron lifetime is $\mntau=885.1\,{\rm s}$.}
\end{figure}

Figure \ref{fig:plot_scenario1} shows the relative difference in
$Y_P$ plotted against \twfo for a neutron lifetime $\mntau=885.1\,{\rm s}$ in
the first scenario.  The relative difference, $\delta Y_P$, is with respect to
the \yp baseline in Eq.\ \eqref{eq:yp_baseline}
\beq\label{eq:deltayp_885}
  \delta Y_P = \frac{Y_P-Y_P^{\rm (base)}[\mntau=885.1\,{\rm s}]}
  {Y_P^{\rm (base)}[\mntau=885.1\,{\rm s}]}.
\eeq
The first scenario is able to induce a change in \yp at the percent level.  The
rates for the capture processes are proportional to $\tcm^5$ for high
temperature, implying that the free neutron decay rate has a vanishingly small
contribution in effecting the \npratio in this regime.  As a result, the
scenario of altered WFO considered here is immutable for $\twfo\gtrsim$ a few
MeV and the curve in Fig.\ \ref{fig:plot_scenario1} begins to asymptote as
\twfo increases into this range of comoving temperature parameter.

The first altered scenario of WFO has no effect for values of \twfo set after
\heiv formation, $\tcm\lesssim50\,{\rm keV}$, and the relative change in \yp is
zero as shown in Fig.\ \ref{fig:plot_scenario1}.  In addition, $\delta\yp$ has
asymptoted for $\tcm\gtrsim1\,{\rm MeV}$.  $\delta\yp$ undergoes a rapid ascent
for the range $400\,{\rm keV}\gtrsim\tcm\gtrsim100\,{\rm keV}$ and connects the high and
low temperature plateaus for the first scenario. This range encompasses the
epoch when $\lambda_{n\,{\rm decay}}$ is roughly the same order of magnitude as
$\lambda_{\nue n}$ and $\lambda_{e^+n}$ according to Fig.\
\ref{fig:plot_rates_tcm}.  Furthermore, the steep slope of $\delta\yp$ as a
function of \twfo gives us information on {\it when} the \npratio, and by
extension \yp, is the most mutable to the free neutron decay rate.  In this
first scenario of altered WFO, we have abruptly changed the free neutron decay
rate to the vacuum value.  Over the entire history of the early universe, the
free neutron decay rate only changes by a factor of 4, larger than the change
imposed by the first scenario.  Nevertheless, the small temperature range
$400\,{\rm keV}\gtrsim\tcm\gtrsim100,{\rm keV}$, which is $\sim15$ Hubble
times, has an impact on WFO at the percent level in the first scenario.

\subsection{Second scenario}

Figure \ref{fig:plot_rates_tcm} shows that at $\tcm\sim1\,{\rm MeV}$, the
neutron destruction rates ($\lambda_{\nue n}$, $\lambda_{\nue n}$) are roughly
an order of magnitude larger than the neutron creation rates ($\lambda_{e^-p}$,
$\lambda_{\bnue p}$).  Both of the neutron creation rates require a threshold
energy for the incident lepton, reducing the amount of accessible phase space
and lowering the value of the rates compared to the neutron destruction rates.
In the second scenario of altered WFO, we will neglect the processes which
require a threshold for $\tcm<\twfo$.  This is equivalent to an instantaneous
WFO epoch for the two lepton capture rates on protons, Eqs.\
\eqref{eq:urcan1} and \eqref{eq:urcan2}.

Our procedure for this scenario is similar to that of the first.  We calculate
all six \nprates at high temperature.  Once \tcm is lower than \twfo, we
neglect the lepton capture rates on protons in Eqs.\ \eqref{eq:urcan1} and
\eqref{eq:urcan2}.  The expressions implemented in the code are
\begin{align}
  \lambda_i^{\rm (Sc.\,2)}(\tcm)=&
  \begin{cases}
    \lambda_i&{\rm if}\quad\tcm>\twfo\\
    0&{\rm if}\quad\tcm<\twfo
  \end{cases},\nonumber\\
  \lambda_j^{\rm (Sc.\,2)}(\tcm)=&\,\lambda_j,\label{eq:lamsc2}
\end{align}
where $\lambda_i$ are the expressions for the capture rates on protons in Eqs.\
\eqref{eq:urcan1} and \eqref{eq:urcan2}.  The other four \nprates use the
expressions $\lambda_j$ in Eqs.\ \eqref{eq:urcap1}, \eqref{eq:urcap2},
\eqref{eq:ndecay}, and \eqref{eq:indecay} at all temperatures.

\begin{figure}
  \includegraphics[width=\columnwidth]{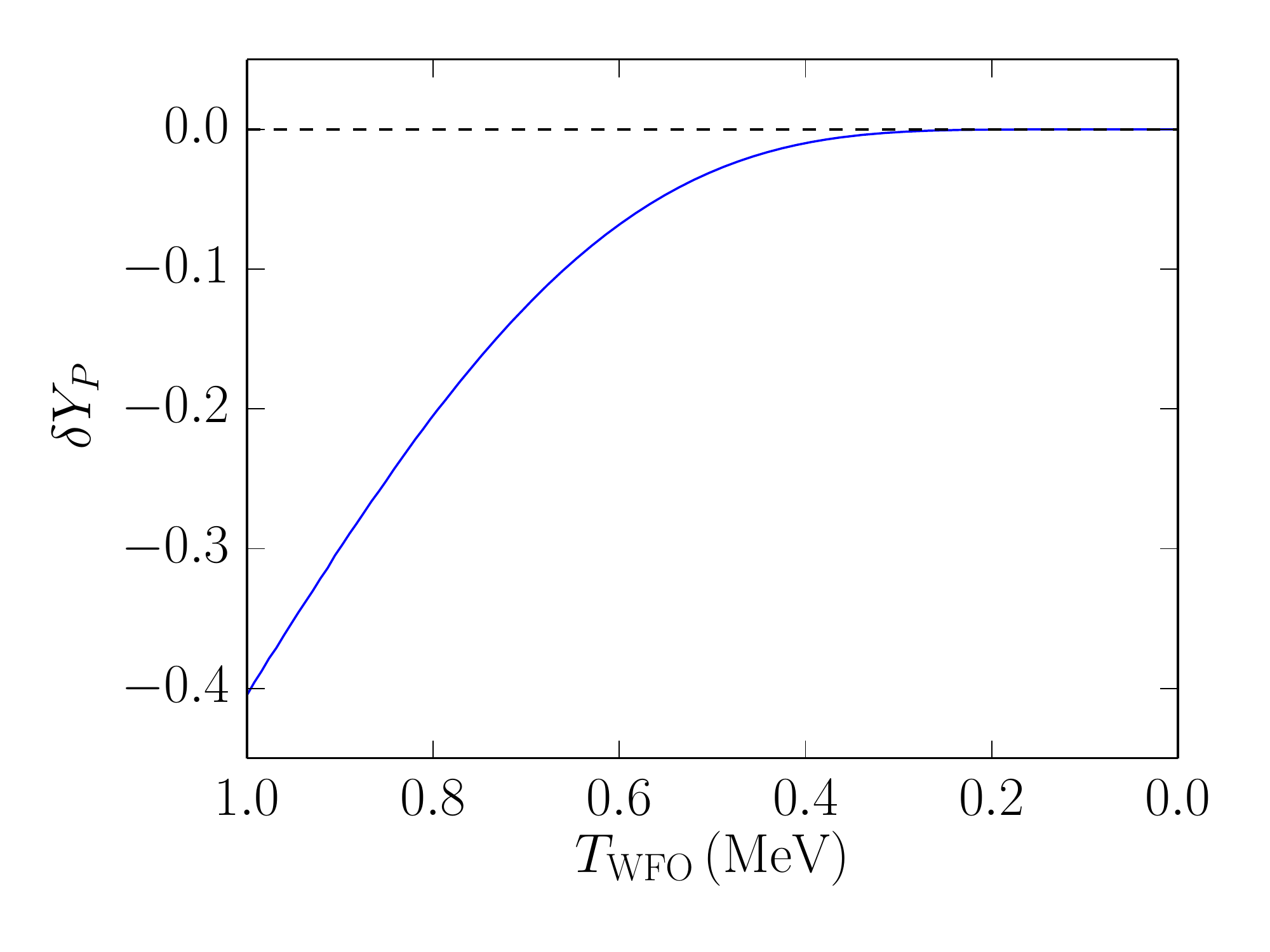}
  \caption{\label{fig:plot_deltayp_twfo_sc2}The relative difference in $Y_P$
  compared to the baseline [Eq.\ \eqref{eq:deltayp_885}] as a function of \twfo
  in the second scenario of altered WFO. All six processes utilize the expressions
  in Eqs.\ \eqref{eq:urcap1} -- \eqref{eq:indecay} until \tcm reaches the
  user-specified \twfo.  For $\tcm<\twfo$, the capture rates on protons, namely
  $e^-p\rightarrow n+\nue$ and $\bnue p\rightarrow ne^+$, are set to zero.  The
  neutron lifetime is $\mntau=885.1\,{\rm s}$.}
\end{figure}

Figure \ref{fig:plot_deltayp_twfo_sc2} shows the results of the second
scenario.  The neutron lifetime is fixed at $\mntau=885.1\,{\rm s}$.  The
horizontal axis gives \twfo and the vertical axis gives the relative change in
$Y_P$ from the baseline in Eq.\ \eqref{eq:yp_baseline}.  If we neglect the
capture rates on protons at $\twfo\sim1\,{\rm MeV}$, we find a $\sim40\%$
decrease in $Y_P$.  The relative difference in \yp decreases with decreasing
\twfo. At $\twfo\sim400\,{\rm keV}$, there exists a $\sim1\%$ change in $Y_P$.
This corresponds to an epoch where the capture rates on protons are roughly two
orders of magnitude smaller than the capture rates on neutrons.  The average
energy of neutrinos at this epoch is $\sim1.2\,{\rm MeV}$, only two-thirds of
the value of $\deltamnp + m_e$.  Only neutrinos in the tail of the distribution
have enough energy to contribute to process \eqref{eq:pn2}.  The exclusion of
process \eqref{eq:pn2}, and to a greater extent process \eqref{eq:pn1}, leads
to a noticeable change well below our initial estimate for WFO, $\tcm\sim\teq$.

\subsection{Third scenario}

In Fig.\ \ref{fig:plot_rates_tcm}, $\lambda_{\nue n}\sim H$ for
$\tcm\sim1.0\,{\rm MeV}$, in line with our earlier estimate for \teq.  For
$\tcm<\teq$, the Hubble expansion rate is larger than any of the \nprates until
$\tcm\simeq50\,{\rm keV}$, where $\lambda_{n\,{\rm decay}}$ becomes larger than
$H$.  In the third scenario of altered WFO, we will implement an even more extreme
instantaneous WFO as compared to the second scenario.

At high temperature we calculate all six of the \nprates using the expressions
in Eqs.\ \eqref{eq:urcap1} -- \eqref{eq:indecay}.  Once \tcm reaches \twfo we
no longer calculate the rates associated with the lepton capture processes on
protons {\it and} neutrons, i.e.\ processes \eqref{eq:np1}, \eqref{eq:np2},
\eqref{eq:pn1}, and \eqref{eq:pn2}, thereby instituting an instantaneous
freeze-out of those specific four lepton capture reactions.  We will calculate
and include the in-medium free neutron decay rate and associated inverse rate
for all temperatures in this scenario.  The expression implemented in the code
is
\begin{align}
  \lambda_i^{\rm (Sc.\,3)}(\tcm)=&
  \begin{cases}
    \lambda_i&{\rm if}\quad\tcm>\twfo\\
    0&{\rm if}\quad\tcm<\twfo
  \end{cases},\nonumber\\
  \lambda_j^{\rm (Sc.\,3)}(\tcm)=&\,\lambda_j,\label{eq:lamsc3}
\end{align}
where $\lambda_i$ are the expressions for the capture rates in Eqs.\
\eqref{eq:urcap1}, \eqref{eq:urcap2}, \eqref{eq:urcan1}, and \eqref{eq:urcan2};
and $\lambda_j$ are the expressions in Eqs.\ \eqref{eq:ndecay} and
\eqref{eq:indecay}.

\begin{figure}
  \begin{center}
     \includegraphics[width=\columnwidth]{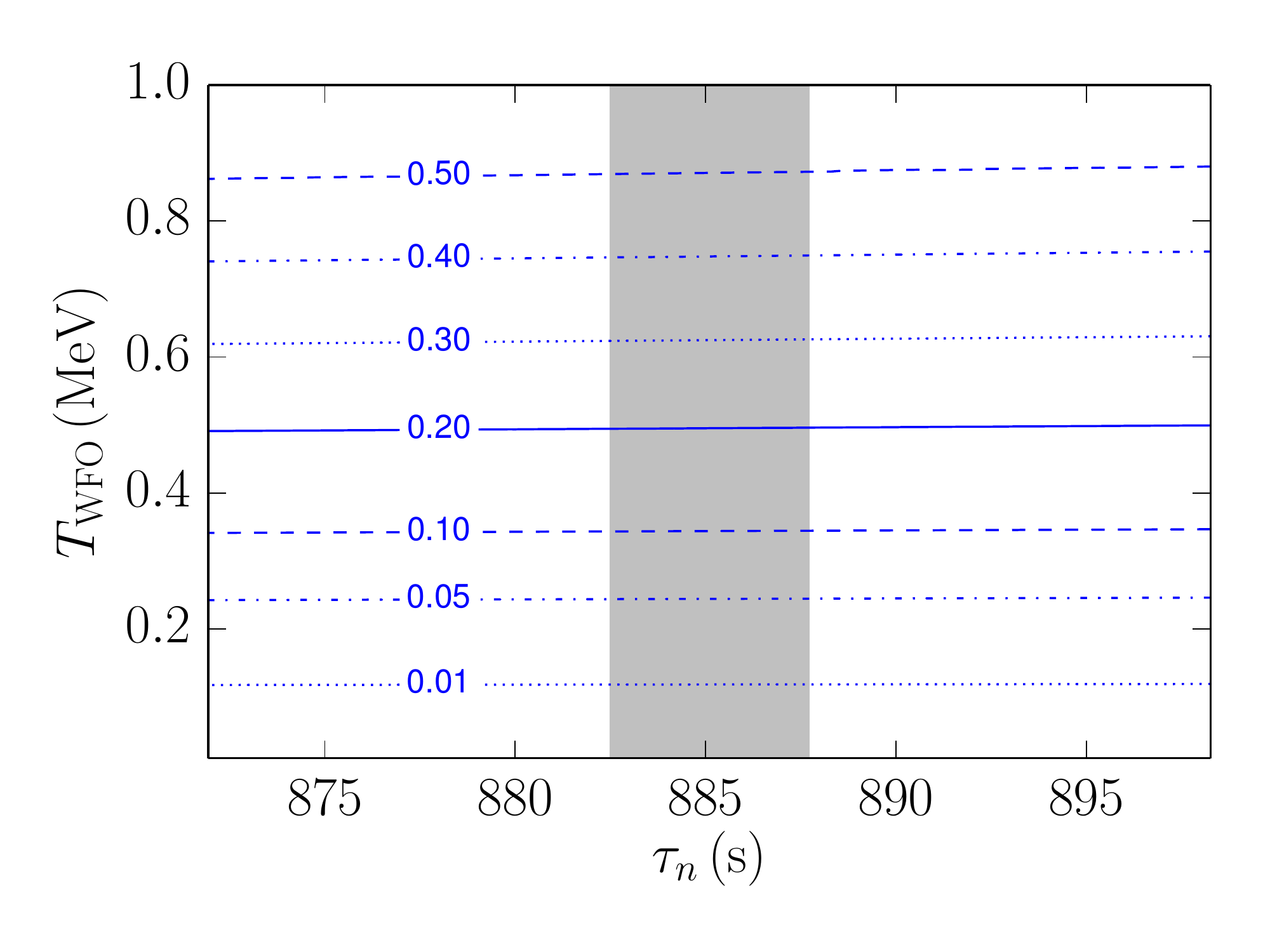}
     \includegraphics[width=\columnwidth]{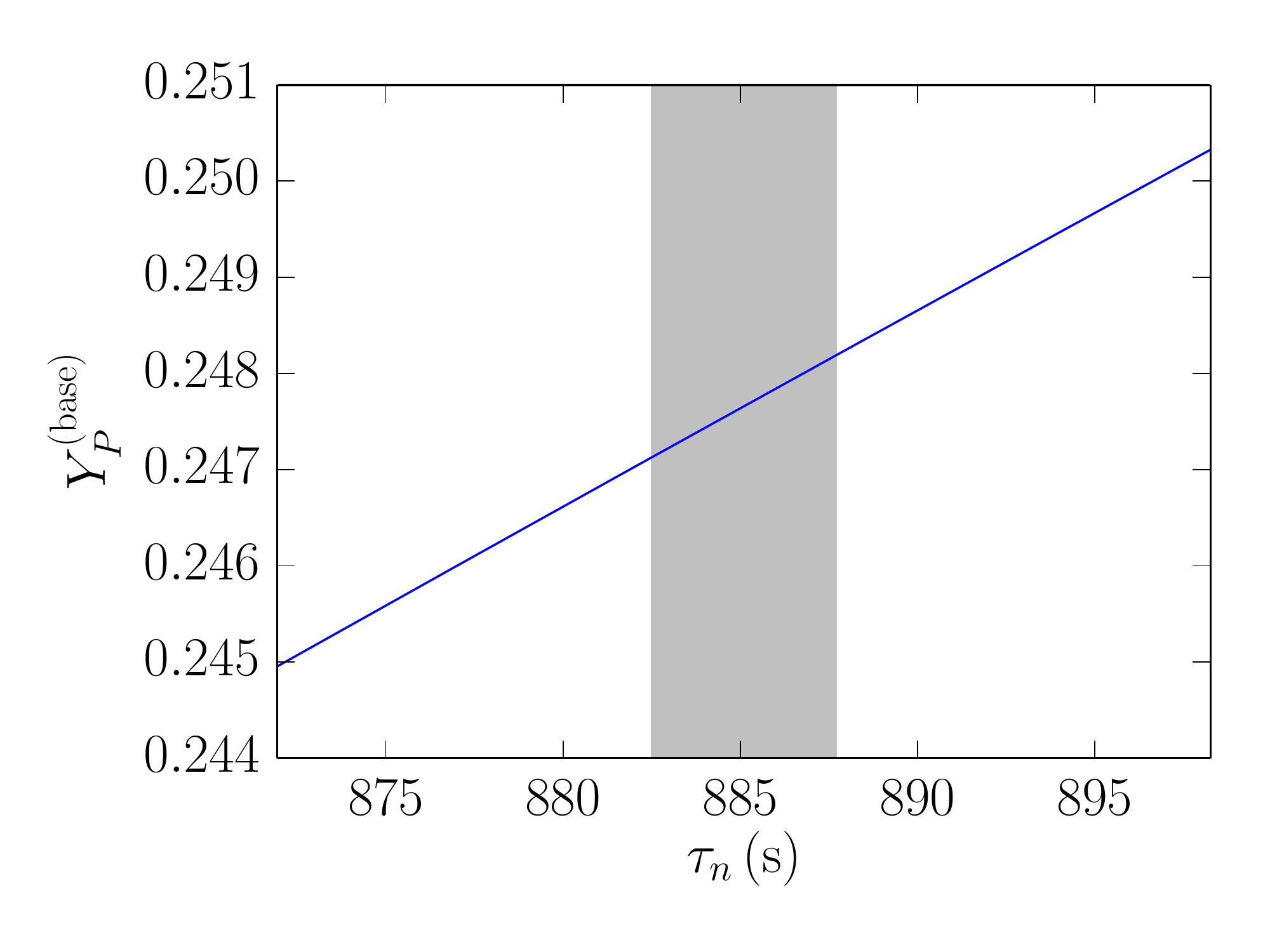}
  \end{center}
  \caption{\label{fig:contour}(Top) \twfo plotted against \mntau for contours
  of constant $\delta Y_P$ in the third scenario of altered WFO.  Here, the
  relative difference in $Y_P$ is given in Eq.\ \eqref{eq:deltayp_mntau}, where
  the baseline value of \yp changes with \mntau.  All six processes utilize the
  expressions in Eqs.\ \eqref{eq:urcap1} -- \eqref{eq:indecay} until \tcm reaches
  the user-specified \twfo.  For $\tcm<\twfo$, the four capture rates, namely
  $\nue+n\leftrightarrow e^-p$ and $e^+n\leftrightarrow\bnue p$, are set to zero.
  (Bottom) The baseline values of $Y_P$ for a given \mntau.  For the baseline, we
  set $\twfo=0$.  In both panels the shaded silver vertical region is the
  one-sigma estimate for $\mntau=885.1\pm2.6\,{\rm s}$ calculated
  from Refs.\ \cite{2012PhRvC..85f5503S,2013PhRvL.111v2501Y}}
\end{figure}

Figure \ref{fig:contour} gives the results of the WFO approximation in the
third scenario.  In this scenario, we define the relative change in $Y_P$ as
the following
\beq\label{eq:deltayp_mntau}
  \delta Y_P = \frac{Y_P-Y_P^{\rm (base)}[\mntau]}{Y_P^{\rm (base)}[\mntau]},
\eeq
where $Y_P^{\rm (base)}[\mntau]$ is the baseline value of $Y_P$ at a given
\mntau if we set $\twfo=0$.  The top panel of Fig.\ \ref{fig:contour} shows
contours of constant $\delta Y_P$ in the \twfo vs.\ \mntau plane.  The bottom
panel shows the baseline values, $Y_P^{\rm (base)}$.  The contours in the top
panel are roughly horizontal, indicating that the relative changes in $Y_P$ are
independent of \mntau.  Our estimate for \teq was 0.9 MeV.  If we set
$\twfo=\teq$, we see that our calculation for $Y_P$ induces a $\sim 50\%$
increase over the baseline.  When calculating $\delta\yp$ for the second
scenario, we found a comparable $35\%$ {\it decrease} in \yp from the baseline
when $\twfo=\teq$ in Fig.\ \ref{fig:plot_deltayp_twfo_sc2}.  The exclusion of
the capture rates on neutrons in going from scenario 2 to scenario 3 results in
a nearly $100\%$ shift in \yp.  In Fig.\ \ref{fig:plot_rates_tcm}, the capture
rates are four orders of magnitude larger than the free-neutron decay rate when
$\tcm=\teq$.  Neglecting the capture rates leads to a significantly different
history of WFO.  The instantaneous WFO approximation improves if we lower
\twfo.  However, even at $\twfo=100\,{\rm keV}$, a $\sim 1\%$ relative
difference in $Y_P$ persists.

\begin{figure}
  \includegraphics[width=\columnwidth]{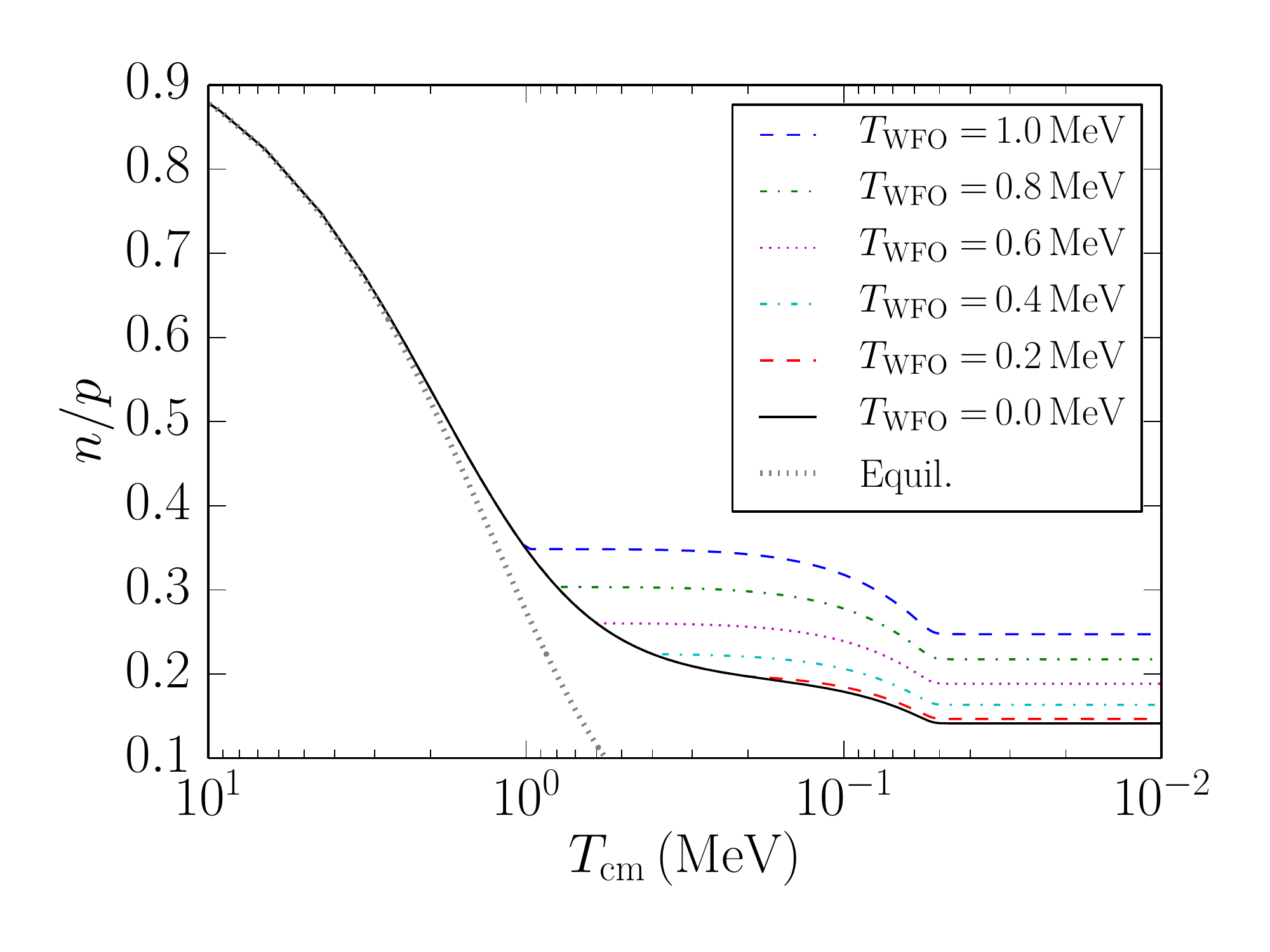}
  \caption{\label{fig:plot_np_tcm}The \npratio as a function of comoving
  temperature parameter \tcm in the third scenario of altered WFO, the same as in
  Fig.\ \ref{fig:contour}.  Each \twfo curve represents a different value for
  temperature $\tcm=\twfo$ where processes \eqref{eq:np1}, \eqref{eq:np2},
  \eqref{eq:pn1}, and \eqref{eq:pn2} are sharply turned off.  All curves use
  $\mntau=885.1\,{\rm s}$.  Also included is a curve (Equil.) assuming the
  \npratio stays in weak equilibrium, i.e.\ $n/p=e^{-\deltamnp/T}$, where we
  neglect the electron chemical potential and set the neutrino chemical potential
  to zero.}
\end{figure}

Figure \ref{fig:plot_np_tcm} shows the evolution of the \npratio as a function
of \tcm for different values of \twfo within the third scenario.  A curve
assuming the \npratio stays in weak equilibrium is included for comparison.
For each nonzero \twfo, the \npratio starkly diverges from the baseline curve
at that value of \twfo.  The value of the \npratio is a monotonic function of
\twfo, approaching the baseline for decreasing values of \twfo.  The evolution
of the \npratio abruptly ceases at $\tcm\sim50\,{\rm keV}$ - the conclusion of
\heiv formation.

The value of the \npratio is important for BBN.  Before BBN commences, the
abundances are in nuclear statistical equilibrium.  For a nucleus with
mass number $A$ and atomic number $Z$, the NSE abundance is \cite{B2FH}
\beq
  Y_X^{\rm (NSE)}\simeq\,Y_p^ZY_n^{A-Z}
  2^{(A-3)/2}\pi^{3(A-1)/2}
  g_XA^{3/2}\label{eq:nse}
  \left[\frac{n_b}{(Tm_b)^{3/2}}\right]^{A-1}e^{B_X/T},
\eeq
In Eq.\ \eqref{eq:nse}, $g_X$ is the spin and $B_X$ is the binding energy of
nucleus $X$.  $m_b$ is the baryon rest mass energy.  We can relate the baryon
number density to the entropy per baryon in the plasma
\beq
  \spl=\frac{1}{n_b}\frac{2\pi^2}{45}\gstars T^3,
\eeq
where \gstars is the statistical entropic weight \cite{1990eaun.book.....K}.
\spl is nearly constant during WFO (see Ref.\ \cite{transport_paper} for a
discussion of entropy flows during the weak decoupling, WFO, and BBN epochs).
Eq.\ \eqref{eq:nse} shows that $Y_X$ depends on time/temperature through the
quantities $Y_p^Z$, $Y_n^{A-Z}$, and $T^{9(1-A)/2}$ before the abundance goes
out of equilibrium.  As a result, $Y_X$ will depart from NSE at specific values
of $Y_p$ and $Y_n$, or equivalently, at a specific value of the \npratio.  The
evolution of $Y_X$ then proceeds to follow from a Boltzmann equation.  The
Boltzmann equation is sensitive to the initial conditions, which include the
\npratio in this case.

\begin{figure}
  \includegraphics[width=\columnwidth]{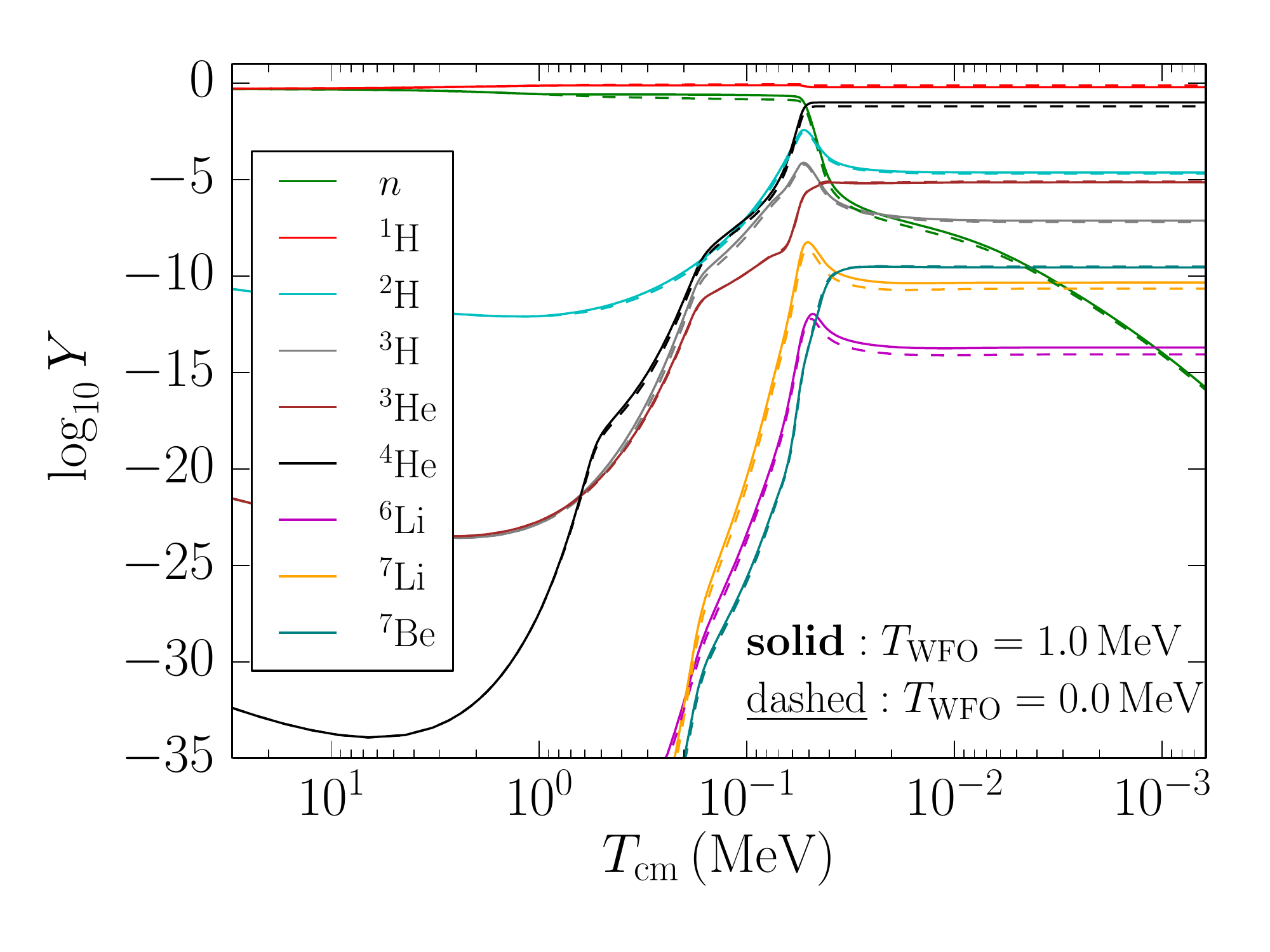}
  \caption{\label{fig:plot_abunds_tcm}The abundances as a function of comoving
  temperature parameter \tcm in the third scenario of altered WFO, the same as in
  Fig.\ \ref{fig:contour}.  The solid lines show the abundances if
  $\twfo=1.0\,{\rm MeV}$.  The dashed lines show the abundances if
  $\twfo=0.0\,{\rm MeV}$, i.e.\ if the \nprates are calculated at all times.  The
  neutron lifetime is $\mntau=885.1\,{\rm s}$.}
\end{figure}

Figure \ref{fig:plot_abunds_tcm} shows the evolution of the primordial
abundances as functions of \tcm.  The dashed curves correspond to the baseline
case, $\twfo=0$, whereas the solid curves correspond to the case
$\twfo=1.0\,{\rm MeV}$ in the third scenario.  Although the vertical axis
ranges over 36 orders of magnitude, the instantaneous WFO approximation induces
a discernible difference on the evolution of the abundances.  Furthermore, the
solid and dashed curves begin diverging for $\tcm\lesssim1.0\,{\rm MeV}$, the
\twfo value for the solid curves.  \heiv is still in NSE at $\tcm=1.0\,{\rm
MeV}$.  In essence, we have precipitated an earlier epoch of out-of-equilibrium
evolution when we set $\twfo=1.0\,{\rm MeV}$.  Therefore, we have a different
\npratio and initial condition as dictated by Eq.\ \eqref{eq:nse} compared to
the baseline case.  The result is a different evolution for \heiv.  The helium
abundances differ by $\sim50\%$  for the two cases at the conclusion of BBN.
\heiv is indeed sensitive to the \npratio, but to a lesser extent so are
deuterium and $^3{\rm He}$.  For this scenario compared to the baseline,
deuterium differs by $\sim44\%$ and $^3{\rm He}$ differs by $\sim14\%$.

\subsection{Fourth scenario}

For the fourth and last scenario, we combine scenarios 1 and 3.  The expression
for the rates in the code is
\begin{align}
  \lambda_i^{\rm (Sc.\,4)}(\tcm)=&
  \begin{cases}
    \lambda_i&{\rm if}\quad\tcm>\twfo\\
    0&{\rm if}\quad\tcm<\twfo
  \end{cases},\nonumber\\
  \lambda_{n\,{\rm decay}}^{\rm (Sc.\,4)}(\tcm)=&
  \begin{cases}
    \lambda_{n\,{\rm decay}}&{\rm if}\quad\tcm>\twfo\\
    1/\mntau&{\rm if}\quad\tcm<\twfo
  \end{cases},\nonumber\\
  \lambda_{\bnue e^-p}^{\rm (Sc.\,4)}(\tcm)=&\,\lambda_{\bnue e^-p},
\end{align}
where $\lambda_i$ are the expressions for the capture rates in Eqs.\
\eqref{eq:urcap1}, \eqref{eq:urcap2}, \eqref{eq:urcan1}, and \eqref{eq:urcan2}.
In other words, in the fourth scenario we assume a sharp transition at a chosen
\twfo after which we neglect all lepton capture processes and use an unblocked,
vacuum decay rate for free neutrons.

\begin{figure}
  \includegraphics[width=\columnwidth]{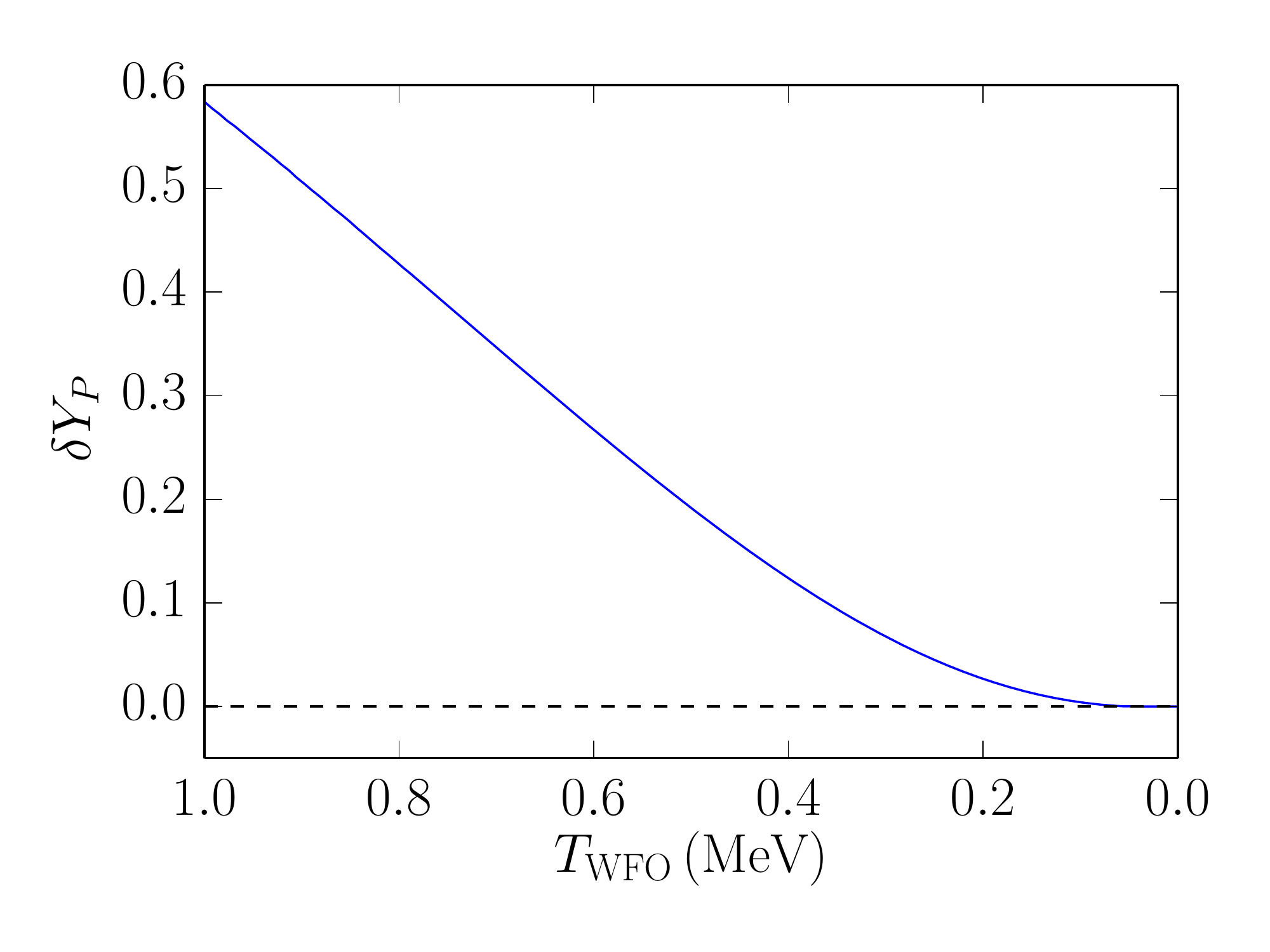}
  \caption{\label{fig:plot_deltayp_twfo_sc4}The relative difference in $Y_P$
  compared to the baseline [Eq.\ \eqref{eq:deltayp_885}] as a function of \twfo
  in the fourth scenario of altered WFO. All six processes utilize the
  expressions in Eqs.\ \eqref{eq:urcap1} -- \eqref{eq:indecay} until \tcm reaches
  the user-specified \twfo.  For $\tcm<\twfo$, the capture rates, namely
  $e^-p\leftrightarrow n+\nue$ and $\bnue p\leftrightarrow ne^+$, are set to
  zero, and the free neutron decay rate is set to the vacuum rate.  Our adopted neutron
  lifetime is $\mntau=885.1\,{\rm s}$.}
\end{figure}

Figure \ref{fig:plot_deltayp_twfo_sc4} shows the relative change in $Y_P$
plotted against \twfo.  Scenario 1 produced a decrease in \yp, and scenario 3
produced an increase in \yp over the baseline.  In addition, scenario 3
produced much larger changes in \yp for equal \twfo.  As a result, $\delta\yp$
for this scenario seen in Fig.\ \ref{fig:plot_deltayp_twfo_sc4} closely follows
that of the third scenario.  If we consider the top panel of Fig.\
\ref{fig:contour}, we see a $\sim60\%$ change in \yp for $\twfo=1.0\,{\rm MeV}$
at the central value $\mntau=885.1\,{\rm s}$.  This is in line with the change
at $\twfo=1.0\,{\rm MeV}$ for scenario 4 in Fig.\
\ref{fig:plot_deltayp_twfo_sc4}.  To be more precise, $\delta\yp$ for scenario
4 is a half percent lower than scenario 3 for $\twfo=1.0\,{\rm MeV}$.  At
$\twfo=1.0\,{\rm MeV}$ in scenario 1, $\yp$ is more than a full percent lower
than the baseline.  The values for $\delta\yp$ in scenario 4 are not perfect
incoherent sums of the values in scenarios 1 and 3, but are also not widely
discrepant.  At lower temperatures, the effect from the blocking factors in
free neutron decay play a lesser role than they do at higher temperatures.  As
a result, scenario 4 becomes equivalent to scenario 3 and $\delta\yp$ assume
the same values for both scenarios at lower temperature.  When
$\twfo\sim100\,{\rm keV}$, we find a $\sim1\%$ change in $Y_P$ for both
scenarios 3 and 4.  Note that if we choose $\twfo =0.7\,{\rm MeV}$, this
scenario yields a \yp nearly $40\%$ above the baseline standard BBN value.
Clearly, Fig.\ \ref{fig:plot_deltayp_twfo_sc4} shows that any choice of a sharp
\twfo in scenario 4 results in incorrect nucleosynthesis.

\section{Conclusion}
\label{sec:conclusion}

We have presented calculations which show that neglecting almost any of the
isospin-changing weak interaction processes in the WFO epoch is unjustified and
results in incorrect BBN.  In particular, the idea that there is a WFO
temperature, \twfo, below which lepton capture on neutrons and protons, and
blocking factors for in-medium free neutron decay can be neglected, is
inaccurate and misleading.  We have quantified how the effect of individual
isospin-changing weak interactions figures into the evolution of the \npratio
and light-element abundances.

Most extant BBN calculations
\cite{GFKP-5pts:2014mn,letsgoeu2,2008CoPhC.178..956P,2012CoPhC.183.1822A}
capture the physics of WFO correctly.  It is nevertheless surprising that
lepton capture reactions, even those with an energy threshold, are important in
calculating final helium and deuterium abundances.  Fig.\
\ref{fig:plot_rates_tcm}  illustrates why this result is surprising.  Note that
in Fig.\ \ref{fig:plot_rates_tcm}, the Hubble expansion rate dominates over all
lepton capture isospin-changing reactions for $\tcm<0.9$ MeV, but the
comparison between individual reaction rates and the Hubble expansion rate can
be misleading.  A more revealing procedure is to solve for the solution of the
\npratio, tantamount to a solution of the two coupled first-order differential
equations, \eqref{eq:dyndt} and \eqref{eq:dypdt}.  These solutions involve
exponentials of integrals of the Hubble expansion rate convolved with
isospin-changing weak interaction rates, as shown in Ref.\
\cite{1989RvMP...61...25B} for the early universe and Ref.\
\cite{1993PhRvL..71.1965Q} for the very similar situation in a neutrino-driven
wind around a hot proto-neutron star.  In essence, these solutions show that
the isospin-changing weak interactions have an integrated, cumulative effect
which must be accounted for.  Although these weak interaction rates drop
dramatically as the temperature drops, the Hubble expansion rate also drops -
necessitating a numerical approach in integrating these rates.

Future prospects for high-precision determinations of the primordial abundances
of helium, e.g., from cosmic microwave background stage 4 experiments
\cite{cmbs4uofc}, and deuterium, from thirty-meter-class telescopes
\cite{Cooke:2014do,2016MNRAS.455.1512C}, heighten the prospects for BBN
constraints on and probes of BSM physics in the early universe.  Our results
point up the sensitivity of BBN to slow, subdominant weak interaction processes
at quite low temperatures, $T\sim100\,{\rm keV}$.  In turn, this suggests that
high precision measurements of light-element abundances can translate into
constraints on the thermal physics and any neutrino sector or BSM physics that
affects neutrino or charged-lepton distribution functions and number densities
at this epoch.  For example, neutrino oscillations
\cite{neff:3.046,2016arXiv160606986D}, sterile neutrinos \cite{ABFW_2005_PRD},
particle decay
\cite{FKK:2011di,DarkRadScherrer:2012,2014PhRvD..90h3519I,2014PhRvD..90h3519I,
2014NuPhB.888..248B,2014PhRvD..90j5024B,2015PhRvD..91f3502L}, and large
neutrino magnetic moments \cite{Vassh_tdecoup}, all can produce changes in
active neutrino distributions, number densities, and overall energy density
which can slightly or substantially affect the weak isospin-changing
interactions we consider in this paper.  The weak interaction physics we have
considered in this paper suggests that BBN will be an even more sensitive probe
of these and other BSM physics issues in the future.

\section*{Acknowledgments}

We thank Lauren Gilbert for useful conversations.  We acknowledge the
Integrated Computing Network at Los Alamos National Laboratory for
supercomputer time.  This work was supported in part by NSF grant PHY-1307372
at UC San Diego.


\bibliographystyle{elsarticle-num}
\bibliography{master}

\begin{thebibliography}{10}
\expandafter\ifx\csname url\endcsname\relax
  \def\url#1{\texttt{#1}}\fi
\expandafter\ifx\csname urlprefix\endcsname\relax\def\urlprefix{URL }\fi
\expandafter\ifx\csname href\endcsname\relax
  \def\href#1#2{#2} \def\path#1{#1}\fi

\bibitem{Cooke:2014do}
R.~J. Cooke, M.~Pettini, R.~A. Jorgenson, M.~T. Murphy, C.~C. Steidel,
  Precision measures of the primordial abundance of deuterium, The
  Astrophysical Journal 781~(1) (2014) 31.

\bibitem{2016MNRAS.455.1512C}
R.~{Cooke}, M.~{Pettini}, {The primordial abundance of deuterium: ionization
  correction}, \mnras 455 (2016) 1512--1521.
\newblock \href {http://arxiv.org/abs/1510.03867} {\path{arXiv:1510.03867}},
  \href {http://dx.doi.org/10.1093/mnras/stv2343}
  {\path{doi:10.1093/mnras/stv2343}}.

\bibitem{cmbs4uofc}
D.~Green, Cosmology with {CMB S4} workshop,
  \url{https://cosmo.uchicago.edu/CMB-S4workshops/index.php/UMICH-2015:Neutrino_and_Light_Relativisic_Species_Summary},
  {Accessed}: 2015-11-11.

\bibitem{1998RvMP...70..303S}
D.~N. {Schramm}, M.~S. {Turner}, {Big-bang nucleosynthesis enters the precision
  era}, Reviews of Modern Physics 70 (1998) 303--318.
\newblock \href {http://arxiv.org/abs/astro-ph/9706069}
  {\path{arXiv:astro-ph/9706069}}, \href
  {http://dx.doi.org/10.1103/RevModPhys.70.303}
  {\path{doi:10.1103/RevModPhys.70.303}}.

\bibitem{2002PhR...370..333D}
A.~D. {Dolgov}, {Neutrinos in cosmology}, \physrep 370 (2002) 333--535.
\newblock \href {http://arxiv.org/abs/hep-ph/0202122}
  {\path{arXiv:hep-ph/0202122}}, \href
  {http://dx.doi.org/10.1016/S0370-1573(02)00139-4}
  {\path{doi:10.1016/S0370-1573(02)00139-4}}.

\bibitem{2012arXiv1208.0032S}
G.~{Steigman}, {Neutrinos And Big Bang Nucleosynthesis}, ArXiv e-prints\href
  {http://arxiv.org/abs/1208.0032} {\path{arXiv:1208.0032}}.

\bibitem{2016RvMP...88a5004C}
R.~H. {Cyburt}, B.~D. {Fields}, K.~A. {Olive}, T.-H. {Yeh}, {Big bang
  nucleosynthesis: Present status}, Reviews of Modern Physics 88~(1) (2016)
  015004.
\newblock \href {http://arxiv.org/abs/1505.01076} {\path{arXiv:1505.01076}},
  \href {http://dx.doi.org/10.1103/RevModPhys.88.015004}
  {\path{doi:10.1103/RevModPhys.88.015004}}.

\bibitem{B2FH}
E.~M. {Burbidge}, G.~R. {Burbidge}, W.~A. {Fowler}, F.~{Hoyle}, {Synthesis of
  the Elements in Stars}, Reviews of Modern Physics 29 (1957) 547--650.
\newblock \href {http://dx.doi.org/10.1103/RevModPhys.29.547}
  {\path{doi:10.1103/RevModPhys.29.547}}.

\bibitem{Dolgov:1997ne}
A.~D. {Dolgov}, S.~H. {Hansen}, D.~V. {Semikoz}, {Non-equilibrium corrections
  to the spectra of massless neutrinos in the early universe}, Nuclear Physics
  B 503 (1997) 426--444.
\newblock \href {http://arxiv.org/abs/hep-ph/9703315}
  {\path{arXiv:hep-ph/9703315}}, \href
  {http://dx.doi.org/10.1016/S0550-3213(97)00479-3}
  {\path{doi:10.1016/S0550-3213(97)00479-3}}.

\bibitem{neff:3.046}
G.~{Mangano}, G.~{Miele}, S.~{Pastor}, T.~{Pinto}, O.~{Pisanti}, P.~D.
  {Serpico}, {Relic neutrino decoupling including flavour oscillations},
  Nuclear Physics B 729 (2005) 221--234.
\newblock \href {http://arxiv.org/abs/hep-ph/0506164}
  {\path{arXiv:hep-ph/0506164}}, \href
  {http://dx.doi.org/10.1016/j.nuclphysb.2005.09.041}
  {\path{doi:10.1016/j.nuclphysb.2005.09.041}}.

\bibitem{transport_paper}
E.~{Grohs}, G.~M. {Fuller}, C.~T. {Kishimoto}, M.~W. {Paris}, A.~{Vlasenko},
  {Neutrino energy transport in weak decoupling and big bang nucleosynthesis},
  \prd 93~(8) (2016) 083522.
\newblock \href {http://arxiv.org/abs/1512.02205} {\path{arXiv:1512.02205}},
  \href {http://dx.doi.org/10.1103/PhysRevD.93.083522}
  {\path{doi:10.1103/PhysRevD.93.083522}}.

\bibitem{Wagoner:1966pv}
R.~V. Wagoner, W.~A. Fowler, F.~Hoyle, {On the Synthesis of elements at very
  high temperatures}, Astrophys.J. 148 (1967) 3--49.
\newblock \href {http://dx.doi.org/10.1086/149126} {\path{doi:10.1086/149126}}.

\bibitem{Wagoner:1969sy}
R.~V. {Wagoner}, {Synthesis of the Elements Within Objects Exploding from Very
  High Temperatures}, \apjs 18 (1969) 247.
\newblock \href {http://dx.doi.org/10.1086/190191} {\path{doi:10.1086/190191}}.

\bibitem{2010PhRvD..82l5017F}
G.~M. {Fuller}, C.~J. {Smith}, {Nuclear weak interaction rates in primordial
  nucleosynthesis}, \prd 82~(12) (2010) 125017.
\newblock \href {http://arxiv.org/abs/1009.0277} {\path{arXiv:1009.0277}},
  \href {http://dx.doi.org/10.1103/PhysRevD.82.125017}
  {\path{doi:10.1103/PhysRevD.82.125017}}.

\bibitem{GFKP-5pts:2014mn}
E.~{Grohs}, G.~M. {Fuller}, C.~T. {Kishimoto}, M.~W. {Paris}, {Probing neutrino
  physics with a self-consistent treatment of the weak decoupling,
  nucleosynthesis, and photon decoupling epochs}, \jcap 5 (2015) 17.
\newblock \href {http://arxiv.org/abs/1502.02718} {\path{arXiv:1502.02718}}.

\bibitem{letsgoeu2}
L.~{Kawano}, {Let's go: Early universe 2. Primordial nucleosynthesis the
  computer way}, NASA STI/Recon Technical Report 92 (1992) 25163.

\bibitem{2008CoPhC.178..956P}
O.~{Pisanti}, A.~{Cirillo}, S.~{Esposito}, F.~{Iocco}, G.~{Mangano},
  G.~{Miele}, P.~D. {Serpico}, {PArthENoPE: Public algorithm evaluating the
  nucleosynthesis of primordial elements}, Computer Physics Communications 178
  (2008) 956--971.
\newblock \href {http://arxiv.org/abs/0705.0290} {\path{arXiv:0705.0290}},
  \href {http://dx.doi.org/10.1016/j.cpc.2008.02.015}
  {\path{doi:10.1016/j.cpc.2008.02.015}}.

\bibitem{2012CoPhC.183.1822A}
A.~{Arbey}, {AlterBBN: A program for calculating the BBN abundances of the
  elements in alternative cosmologies}, Computer Physics Communications 183
  (2012) 1822--1831.
\newblock \href {http://arxiv.org/abs/1106.1363} {\path{arXiv:1106.1363}},
  \href {http://dx.doi.org/10.1016/j.cpc.2012.03.018}
  {\path{doi:10.1016/j.cpc.2012.03.018}}.

\bibitem{2016arXiv160509383B}
D.~N. {Blaschke}, V.~{Cirigliano}, {Neutrino Quantum Kinetic Equations: The
  Collision Term}, ArXiv e-prints\href {http://arxiv.org/abs/1605.09383}
  {\path{arXiv:1605.09383}}.

\bibitem{1991NuPhB.349..743B}
R.~{Barbieri}, A.~{Dolgov}, {Neutrino oscillations in the early universe},
  Nuclear Physics B 349 (1991) 743--753.
\newblock \href {http://dx.doi.org/10.1016/0550-3213(91)90396-F}
  {\path{doi:10.1016/0550-3213(91)90396-F}}.

\bibitem{AkhmedovBerezhiani}
E.~K. {Akhmedov}, Z.~G. {Berezhiani}, {Implications of Majorana neutrino
  transition magnetic moments for neutrino signals from supernovae}, Nuclear
  Physics B 373 (1992) 479--497.
\newblock \href {http://dx.doi.org/10.1016/0550-3213(92)90441-D}
  {\path{doi:10.1016/0550-3213(92)90441-D}}.

\bibitem{1993APh.....1..165R}
G.~{Raffelt}, G.~{Sigl}, {Neutrino flavor conversion in a supernova core},
  Astroparticle Physics 1 (1993) 165--183.
\newblock \href {http://arxiv.org/abs/astro-ph/9209005}
  {\path{arXiv:astro-ph/9209005}}, \href
  {http://dx.doi.org/10.1016/0927-6505(93)90020-E}
  {\path{doi:10.1016/0927-6505(93)90020-E}}.

\bibitem{2005PhRvD..71i3004S}
P.~{Strack}, A.~{Burrows}, {Generalized Boltzmann formalism for oscillating
  neutrinos}, \prd 71~(9) (2005) 093004.
\newblock \href {http://arxiv.org/abs/hep-ph/0504035}
  {\path{arXiv:hep-ph/0504035}}, \href
  {http://dx.doi.org/10.1103/PhysRevD.71.093004}
  {\path{doi:10.1103/PhysRevD.71.093004}}.

\bibitem{2007JPhG...34...47B}
A.~B. {Balantekin}, Y.~{Pehlivan}, {Neutrino neutrino interactions and flavour
  mixing in dense matter}, Journal of Physics G Nuclear Physics 34 (2007)
  47--65.
\newblock \href {http://arxiv.org/abs/astro-ph/0607527}
  {\path{arXiv:astro-ph/0607527}}, \href
  {http://dx.doi.org/10.1088/0954-3899/34/1/004}
  {\path{doi:10.1088/0954-3899/34/1/004}}.

\bibitem{2013PhRvD..87k3010V}
C.~{Volpe}, D.~{V{\"a}{\"a}n{\"a}nen}, C.~{Espinoza}, {Extended evolution
  equations for neutrino propagation in astrophysical and cosmological
  environments}, \prd 87~(11) (2013) 113010.
\newblock \href {http://arxiv.org/abs/1302.2374} {\path{arXiv:1302.2374}},
  \href {http://dx.doi.org/10.1103/PhysRevD.87.113010}
  {\path{doi:10.1103/PhysRevD.87.113010}}.

\bibitem{2013PrPNP..71..162B}
A.~B. {Balantekin}, G.~M. {Fuller}, {Neutrinos in cosmology and astrophysics},
  Progress in Particle and Nuclear Physics 71 (2013) 162--166.
\newblock \href {http://arxiv.org/abs/1303.3874} {\path{arXiv:1303.3874}},
  \href {http://dx.doi.org/10.1016/j.ppnp.2013.03.008}
  {\path{doi:10.1016/j.ppnp.2013.03.008}}.

\bibitem{Gouvea}
A.~{de Gouv{\^e}a}, S.~{Shalgar}, {Transition magnetic moments and collective
  neutrino oscillations: three-flavor effects and detectability}, \jcap 4
  (2013) 18.
\newblock \href {http://arxiv.org/abs/1301.5637} {\path{arXiv:1301.5637}},
  \href {http://dx.doi.org/10.1088/1475-7516/2013/04/018}
  {\path{doi:10.1088/1475-7516/2013/04/018}}.

\bibitem{2014PhRvD..90l5040S}
J.~{Serreau}, C.~{Volpe}, {Neutrino-antineutrino correlations in dense
  anisotropic media}, \prd 90~(12) (2014) 125040.
\newblock \href {http://arxiv.org/abs/1409.3591} {\path{arXiv:1409.3591}},
  \href {http://dx.doi.org/10.1103/PhysRevD.90.125040}
  {\path{doi:10.1103/PhysRevD.90.125040}}.

\bibitem{2015PhLB..747...27C}
V.~{Cirigliano}, G.~M. {Fuller}, A.~{Vlasenko}, {A new spin on neutrino quantum
  kinetics}, Physics Letters B 747 (2015) 27--35.
\newblock \href {http://arxiv.org/abs/1406.5558} {\path{arXiv:1406.5558}},
  \href {http://dx.doi.org/10.1016/j.physletb.2015.04.066}
  {\path{doi:10.1016/j.physletb.2015.04.066}}.

\bibitem{Dodelson:2003mc}
S.~Dodelson, \href{http://books.google.com/books?id=3oPRxdXJexcC}{Modern
  Cosmology}, Academic Press, Academic Press, 2003.
\newline\urlprefix\url{http://books.google.com/books?id=3oPRxdXJexcC}

\bibitem{1990eaun.book.....K}
E.~W. {Kolb}, M.~S. {Turner}, {The Early Universe.}, Addison-Wesley Publishing
  Co., 1990.

\bibitem{1982PhRvD..26.2694D}
D.~A. {Dicus}, E.~W. {Kolb}, A.~M. {Gleeson}, E.~C.~G. {Sudarshan}, V.~L.
  {Teplitz}, M.~S. {Turner}, {Primordial nucleosynthesis including radiative,
  Coulomb, and finite-temperature corrections to weak rates}, \prd 26 (1982)
  2694--2706.
\newblock \href {http://dx.doi.org/10.1103/PhysRevD.26.2694}
  {\path{doi:10.1103/PhysRevD.26.2694}}.

\bibitem{1982NuPhB.209..372C}
J.-L. {Cambier}, J.~R. {Primack}, M.~{Sher}, {Finite temperature radiative
  corrections to neutron decay and related processes}, Nucl. Phys. B 209 (1982)
  372--388.
\newblock \href {http://dx.doi.org/10.1016/0550-3213(82)90262-0}
  {\path{doi:10.1016/0550-3213(82)90262-0}}.

\bibitem{1999PhRvD..59j3502L}
R.~E. {Lopez}, M.~S. {Turner}, {Precision prediction for the big-bang abundance
  of primordial $^{4}$He}, \prd 59~(10) (1999) 103502.
\newblock \href {http://arxiv.org/abs/astro-ph/9807279}
  {\path{arXiv:astro-ph/9807279}}, \href
  {http://dx.doi.org/10.1103/PhysRevD.59.103502}
  {\path{doi:10.1103/PhysRevD.59.103502}}.

\bibitem{1980ApJS...42..447F}
G.~M. {Fuller}, W.~A. {Fowler}, M.~J. {Newman}, {Stellar weak-interaction rates
  for sd-shell nuclei. I - Nuclear matrix element systematics with application
  to Al-26 and selected nuclei of importance to the supernova problem}, \apjs
  42 (1980) 447--473.
\newblock \href {http://dx.doi.org/10.1086/190657} {\path{doi:10.1086/190657}}.

\bibitem{2010PhRvD..81f5027S}
C.~J. {Smith}, G.~M. {Fuller}, {Weak interaction rate Coulomb corrections in
  big bang nucleosynthesis}, \prd 81~(6) (2010) 065027.
\newblock \href {http://arxiv.org/abs/0905.2781} {\path{arXiv:0905.2781}},
  \href {http://dx.doi.org/10.1103/PhysRevD.81.065027}
  {\path{doi:10.1103/PhysRevD.81.065027}}.

\bibitem{1997PhRvD..56.3191L}
R.~E. {Lopez}, M.~S. {Turner}, G.~{Gyuk}, {Effect of finite nucleon mass on
  primordial nucleosynthesis}, \prd 56 (1997) 3191--3197.
\newblock \href {http://arxiv.org/abs/astro-ph/9703065}
  {\path{arXiv:astro-ph/9703065}}, \href
  {http://dx.doi.org/10.1103/PhysRevD.56.3191}
  {\path{doi:10.1103/PhysRevD.56.3191}}.

\bibitem{2009PhRvD..79j5001S}
C.~J. {Smith}, G.~M. {Fuller}, M.~S. {Smith}, {Big bang nucleosynthesis with
  independent neutrino distribution functions}, \prd 79~(10) (2009) 105001.
\newblock \href {http://arxiv.org/abs/0812.1253} {\path{arXiv:0812.1253}},
  \href {http://dx.doi.org/10.1103/PhysRevD.79.105001}
  {\path{doi:10.1103/PhysRevD.79.105001}}.

\bibitem{2012PhRvC..85f5503S}
A.~{Steyerl}, J.~M. {Pendlebury}, C.~{Kaufman}, S.~S. {Malik}, A.~M. {Desai},
  {Quasielastic scattering in the interaction of ultracold neutrons with a
  liquid wall and application in a reanalysis of the Mambo I neutron-lifetime
  experiment}, \prc 85~(6) (2012) 065503.
\newblock \href {http://dx.doi.org/10.1103/PhysRevC.85.065503}
  {\path{doi:10.1103/PhysRevC.85.065503}}.

\bibitem{2013PhRvL.111v2501Y}
A.~T. {Yue}, M.~S. {Dewey}, D.~M. {Gilliam}, G.~L. {Greene}, A.~B. {Laptev},
  J.~S. {Nico}, W.~M. {Snow}, F.~E. {Wietfeldt}, {Improved Determination of the
  Neutron Lifetime}, Physical Review Letters 111~(22) (2013) 222501.
\newblock \href {http://arxiv.org/abs/1309.2623} {\path{arXiv:1309.2623}},
  \href {http://dx.doi.org/10.1103/PhysRevLett.111.222501}
  {\path{doi:10.1103/PhysRevLett.111.222501}}.

\bibitem{1989RvMP...61...25B}
J.~{Bernstein}, L.~S. {Brown}, G.~{Feinberg}, {Cosmological helium production
  simplified}, Reviews of Modern Physics 61 (1989) 25--39.
\newblock \href {http://dx.doi.org/10.1103/RevModPhys.61.25}
  {\path{doi:10.1103/RevModPhys.61.25}}.

\bibitem{1993PhRvL..71.1965Q}
Y.-Z. {Qian}, G.~M. {Fuller}, G.~J. {Mathews}, R.~W. {Mayle}, J.~R. {Wilson},
  S.~E. {Woosley}, {Connection between flavor mixing of cosmologically
  significant neutrinos and heavy element nucleosynthesis in supernovae},
  Physical Review Letters 71 (1993) 1965--1968.
\newblock \href {http://dx.doi.org/10.1103/PhysRevLett.71.1965}
  {\path{doi:10.1103/PhysRevLett.71.1965}}.

\bibitem{2016arXiv160606986D}
P.~F. {de Salas}, S.~{Pastor}, {Relic neutrino decoupling with flavour
  oscillations revisited}, ArXiv e-prints\href
  {http://arxiv.org/abs/1606.06986} {\path{arXiv:1606.06986}}.

\bibitem{ABFW_2005_PRD}
K.~{Abazajian}, N.~F. {Bell}, G.~M. {Fuller}, Y.~Y.~Y. {Wong}, {Cosmological
  lepton asymmetry, primordial nucleosynthesis and sterile neutrinos}, \prd
  72~(6) (2005) 063004.
\newblock \href {http://arxiv.org/abs/astro-ph/0410175}
  {\path{arXiv:astro-ph/0410175}}, \href
  {http://dx.doi.org/10.1103/PhysRevD.72.063004}
  {\path{doi:10.1103/PhysRevD.72.063004}}.

\bibitem{FKK:2011di}
G.~M. {Fuller}, C.~T. {Kishimoto}, A.~{Kusenko}, {Heavy sterile neutrinos,
  entropy and relativistic energy production, and the relic neutrino
  background}, ArXiv e-prints\href {http://arxiv.org/abs/1110.6479}
  {\path{arXiv:1110.6479}}.

\bibitem{DarkRadScherrer:2012}
J.~L. {Menestrina}, R.~J. {Scherrer}, {Dark radiation from particle decays
  during big bang nucleosynthesis}, \prd 85~(4) (2012) 047301.
\newblock \href {http://arxiv.org/abs/1111.0605} {\path{arXiv:1111.0605}},
  \href {http://dx.doi.org/10.1103/PhysRevD.85.047301}
  {\path{doi:10.1103/PhysRevD.85.047301}}.

\bibitem{2014PhRvD..90h3519I}
H.~{Ishida}, M.~{Kusakabe}, H.~{Okada}, {Effects of long-lived 10 MeV-scale
  sterile neutrinos on primordial elemental abundances and the effective
  neutrino number}, \prd 90~(8) (2014) 083519.
\newblock \href {http://arxiv.org/abs/1403.5995} {\path{arXiv:1403.5995}},
  \href {http://dx.doi.org/10.1103/PhysRevD.90.083519}
  {\path{doi:10.1103/PhysRevD.90.083519}}.

\bibitem{2014NuPhB.888..248B}
D.~{Boyanovsky}, {Space-time evolution of heavy sterile neutrinos in cascade
  decays}, Nuclear Physics B 888 (2014) 248--270.
\newblock \href {http://arxiv.org/abs/1406.5739} {\path{arXiv:1406.5739}},
  \href {http://dx.doi.org/10.1016/j.nuclphysb.2014.09.018}
  {\path{doi:10.1016/j.nuclphysb.2014.09.018}}.

\bibitem{2014PhRvD..90j5024B}
D.~{Boyanovsky}, {Nearly degenerate heavy sterile neutrinos in cascade decay:
  Mixing and oscillations}, \prd 90~(10) (2014) 105024.
\newblock \href {http://arxiv.org/abs/1409.4265} {\path{arXiv:1409.4265}},
  \href {http://dx.doi.org/10.1103/PhysRevD.90.105024}
  {\path{doi:10.1103/PhysRevD.90.105024}}.

\bibitem{2015PhRvD..91f3502L}
L.~{Lello}, D.~{Boyanovsky}, {Cosmological implications of light sterile
  neutrinos produced after the QCD phase transition}, \prd 91~(6) (2015)
  063502.
\newblock \href {http://arxiv.org/abs/1411.2690} {\path{arXiv:1411.2690}},
  \href {http://dx.doi.org/10.1103/PhysRevD.91.063502}
  {\path{doi:10.1103/PhysRevD.91.063502}}.

\bibitem{Vassh_tdecoup}
N.~{Vassh}, E.~{Grohs}, A.~B. {Balantekin}, G.~M. {Fuller}, {Majorana neutrino
  magnetic moment and neutrino decoupling in big bang nucleosynthesis}, \prd
  92~(12) (2015) 125020.
\newblock \href {http://arxiv.org/abs/1510.00428} {\path{arXiv:1510.00428}},
  \href {http://dx.doi.org/10.1103/PhysRevD.92.125020}
  {\path{doi:10.1103/PhysRevD.92.125020}}.

\end{thebibliography}

\end{document}